\documentclass[]{interact}

\usepackage{lmodern}
\usepackage{epstopdf}
\usepackage{graphicx}
\usepackage[tight]{subfigure}
\usepackage{enumitem}
\usepackage{natbib}
\bibpunct[, ]{(}{)}{;}{a}{}{,}
\renewcommand\bibfont{\fontsize{10}{12}\selectfont}

\usepackage{amsmath,amsfonts,amssymb}
\usepackage{bm}

\usepackage{siunitx}
\usepackage{etoolbox}
\apptocmd{\sloppy}{\hbadness 10000\relax}{}{}   

\DeclareFontFamily{U}{bbm}{}
\DeclareFontShape{U}{bbm}{m}{n}
{  <5> <6> <7> <8> <9> <10> <12> gen * bbm
	<5.5> bbm5 
	<10.95> bbm10%
	<14.4>  bbm12%
	<17.28><20.74><24.88> bbm17}{}

\usepackage{color}

\begin{document}

\title{Revisiting the effect of spatial resolution on information content based on classification results}

\author{
\name{M.~G. Palacio\textsuperscript{a}\thanks{CONTACT M.~G.Palacio. Email: gpalacio@exa.unrc.edu.ar}, S.~B. Ferrero\textsuperscript{a} and A.~C. Frery\textsuperscript{b}}
\affil{\textsuperscript{a}Universidad Nacional de R\'io Cuarto, C\'ordoba, Argentina; \\ \textsuperscript{b}Universidade Federal de Alagoas, Macei\'o, AL,Brazil}
}

\maketitle

\begin{abstract}
	Polarimetric Synthetic Aperture Radar (PolSAR) images are an important source of information.
	Speckle noise gives SAR images a granular appearance that makes interpretation and analysis hard tasks.
	A major issue is the assessment of information content in these kind of images, and how it is affected by usual processing techniques.
    Previous works have resulted in various approaches for quantifying image information content.
    As~\cite{Narayanan02} we study this problem from the classification accuracy viewpoint, 
    focusing in the filtering and the classification stages.
	Thus, through classified images we verify how changing the properties of the input data affects their quality.
	The input is an actual PolSAR image, the control parameters are (i)~the filter (Local Mean or Model Based PolSAR, MBPolSAR) and the size of their support, and (ii)~the classification method (Maximum Likelihood, ML, or Support Vector Machine, SVM), and the output is the precision of the classification algorithm applied to the filtered data.
	To expand the conclusions, this study deals not only with Classification Accuracy, but also with Kappa and Overall Accuracy as measures of map precision.
    Experiments were conducted on two airborne PolSAR images.
    Differently from what was observed by~\cite{Narayanan02}, almost all quality measures are good and increase with degradation, i.e. the filtering algorithms that we used always improve the classification results at least up to supports of size \num{7x7}.
\end{abstract}


\section{Introduction}
\label{sec:intro}

Synthetic Aperture Radar (SAR) systems have been widely employed in a single-channel configuration and have showed their ability to provide high spatial resolution data about the scene. 
The availability of multidimensional SAR systems has made it possible to increase the amount of available information about the surface~\citep{Medasani2017}.

Polarimetric Synthetic Aperture Radar (PolSAR) systems emerged as an important multidimensional configuration that transmit orthogonally polarized pulses towards a target, and record the returned echo. 
Based on the polarization of the incident wave, the object roughness and its physical characteristics, different objects scatter the incident waves in a manner that indicates the nature of the targets~\citep{Helmy2016}.
Therefore PolSAR systems provide the means for a better capture of scene information than univariate SAR~\citep{Frery2014}, allowing the characterization of the targets by means of various channels and the covariance structures among them.
 
The interest on PolSAR data has increased in the last years, mostly after the launch of several spaceborne missions with polarimetric capabilities. 
PolSAR systems provide images of the Earth in both day and night, and for almost all weather conditions. 
Such systems emit an electromagnetic waves with two orthogonal polarizations, horizontal and vertical, and then the receiving antenna collects the backscattered wave in both polarizations
Due to the coherent interference of waves reflected from many elementary scatterers, the return is affected by speckle. 

Although speckle is a true scattering measurement, the complexity of the scattering process makes it necessary to consider it noise~\citep{Foucher2014}.
The alterations produced by speckle can significantly degrade the perceived image quality, complicate its interpretation and analysis, and reduce the precision of target detection and classification~\citep{Ma14}.
Hence, modeling and characterizing speckle, its filtering and assessing its effect on the ability to extract useful information are critical tasks~\citep{Foucher2014}. 

There are plenty of speckle noise reduction techniques available for one-dimensional SAR data \citep{TutorialSpeckleReductionSAR}. 
Instead, for PolSAR data, combating speckle is still a challenging task. 
Any speckle filter has to suppress the noise while preserving the spatial and the polarimetric information \citep{Torres_2014}.

The information content (IC) needs to be quantified before attempting to perform operations to increase it.
However, it is not easily quantifiable; 
the same image may contain different amounts of information depending on the intended application, e.g.,  visualization and classification.

\citet{Narayanan02} used an approach for quantifying the IC based on classification accuracy, and studied how spatial degradation affects the classification of both Landsat Thematic Mapper (TM) and Shuttle Imaging Radar-C (SIR-C) images.
They modeled spatial degradation by convolving with the mean filters and different sizes, and assessed the information content by the accuracy of Maximum Likelihood (ML) classification.
The IC was measured as the number of correctly classified pixels in the whole image, by comparing the classified image with the ground data image.

In previous studies~\citep{Palacio17a}, we followed a similar scheme, using two fully polarimetric images (an Airborne Synthetic Aperture Radar, AIRSAR, and an Uninhabited Aerial Vehicle Synthetic Aperture Radar, UAVSAR).
We applied the Model-Based Polarimetric SAR (MBPolSAR) filter~\citep{Lopez2008} because it considers all the information provided by the covariance matrix, so it improves the estimation of the individual entries and the polarimetric information~\citep{Foucher2014}.
We used a Support Vector Machine (SVM) algorithm to perform the classification stage because it is well suited to handle linearly nonseparable cases~\citep{Burges1998}, which is the situation at hand.
As IC measures we used Kappa, Overall Accuracy, and Classification Accuracy coefficients, derived from the Confusion Matrix obtained from testing data set.
We assessed the method by visual inspection of thematic maps.

Whereas~\cite{Narayanan02} observed a decrease in the IC with respect to spatial degradation, our studies showed the opposite effect. 
To analyze if the change is mainly due to the filter or to the classifier, we propose to assess the performance of four methods resulting from the two filters with the two classifiers.
The main objectives of this article can be summarized as follows: 
(i)~comparing the four methods resulting from the combination of filter and classification; 
(ii)~comparing the performance of the filters (for each classifier); 
(iii)~comparing the performance of the classifiers (for each filter); 
(iv)~evaluating the impact of degradation in each method.

The remainder of this paper is organized as follows. 
In Section~\ref{sec:Metod} we present the filters and the classifiers that we applied and the validation scheme.
Section~\ref{sec:Aplic} shows the application of the proposed methodology to two full PolSAR images and the obtained results. 
Finally, the conclusion and discussion are presented in Section~\ref{sec:Conc}.

\section{METHODOLOGY}
\label{sec:Metod}

To investigate if the change in our previous conclusions, compared with~\cite{Narayanan02}, is due to the filter or to the classifier we propose to assess the performance of four methods resulting from the two filters with the two classifiers.
Table~\ref{table1} shows the proposed methods. 

\begin{table}[htb]
	\tbl{Methods to compare.}
	{\begin{tabular}{lcc} \toprule
			& \multicolumn{2}{c}{Filter} \\ \cmidrule{2-3}
			Classifier & Local Mean (LM) & Model Based PolSAR (MBPolSAR) \\ \midrule
			Maximum Likelihood (ML)&  Method $1$\textsuperscript{$\dagger$} &  Method $2$ \\
			Support Vector Machine (SVM) & Method $3$ & Method $4$\textsuperscript{$\ddagger$} \\ \bottomrule
	\end{tabular}}
	\tabnote{\textsuperscript{$\dagger$}the same as~\cite{Narayanan02}.}
	\tabnote{\textsuperscript{$\ddagger$}the same as~\cite{Palacio17a}.}
	\label{table1}
\end{table}

\subsection{Data filtering}
\label{sec:Filter}

For Methods $1$ and $3$, as in~\cite{Narayanan02}, we first apply an adaptive Refined Lee filter~\citep{Lee1999} of size \num{3x3} and use this image as input to the Local Mean Filter~\citep{Lee1980}.
For Methods $2$ and $4$ we directly use the image as input to the MBPolSAR filter.

While speckle reduction in single-channel SAR is a relatively well-established area; cf., for instance, \cite{TutorialSpeckleReductionSAR},
the speckle reduction problem is more complicated for PolSAR data as 
speckle appears both in the intensity term of each polarization and in the phase terms~\citep{Ma16}. 
In addition, as the PolSAR channels are usually correlated, immediate extensions of the single channel model are not available.
SAR data modeling and SAR data filtering are two different aspects of the same problem, as the availability of a polarimetric noise model would make a correct signal estimation possible~\citep{Lopez2008}.

For every resolution element, a PolSAR system measures the scattering, that can be expressed as the vector
\begin{equation}
\label{s}
\bm s = (S_{hh}, S_{hv}, S_{vv})^T,
\end{equation}
where $\left(.\right)^T$ indicates vector transposition, and $h$ and $v$ denote the horizontal and vertical polarizations, respectively.
\citet{Freeman1998} presented an important description of this three-component representation.

For point scatterers, eq.~\eqref{s} characterizes completely the scattering properties of the target.
Instead, for distributed scatterers, $\bm s$ is random and can be modeled by a multivariate zero-mean complex Gaussian distribution with parameter 
\[
\mathbf{C}=E\left\{\bm s\:\bm s^*\right\},
\] where $\left(.\right)^*$ is the complex conjugate transpose. 

The covariance matrix $\mathbf{C}$ contains all the information that characterizes a distributed scatterer and must be estimated from the recorded data in~\eqref{s}.
The estimation process that is reduced to the estimation of every term of the covariance matrix, is also referred to as PolSAR speckle filtering process.
The ML estimator for $\mathbf{C}$ is 
\begin{equation}
\label{Z1}
\mathbf{Z}=\frac1{n}\sum_{i=1}^{n}\bm s_i \bm s_i^*,
\end{equation}
where $n$ is the number of looks or samples employed to estimate $\mathbf{C}$, and $\bm s_i, i=1,2,\ldots,n$, are realizations of~\eqref{s}.

The estimated covariance matrix $\mathbf{Z}$ is a random variable that has an error with respect to the value to be recovered, $\mathbf{C}$.
This error might be considered as a noise component.
\citet{Lopez2003} have shown that under the hypothesis that $\bm s$ is distributed as a multivariate zero-mean complex Gaussian, speckle noise for PolSAR data is a combination of multiplicative and additive noise components:
\begin{equation}
\label{Z2}
\mathbf{Z}=\mathbf{C}+\mathbf{N}_m+\mathbf{N}_a,
\end{equation}
where $\mathbf{N}_m$ (with zero mean) is the multiplicative speckle noise components for all the elements of the sample covariance matrix, and $\mathbf{N}_a$ (also with zero mean) models the additive speckle noise components which only appears in the off-diagonal elements of the sample covariance matrix.
That is, the diagonal terms of the covariance matrix can be characterized by a multiplicative noise model, whereas the off-diagonal ones have the characteristics of a combined multiplicative and additive noise model~\citep{Lopez2008}. 
Since $E\left\{\mathbf{Z}\right\}=\mathbf{C}$, $\mathbf{Z}$ is an unbiased estimator and therefore the multiplicative-additive model does not involve loss of polarimetric information.

~\cite{Lopez2008} pointed out that, as one may see in~\eqref{Z1}, all the elements of $\mathbf{Z}$ are obtained from two components of $\bm s$, thus the final model for $\mathbf{Z}$ may be derived from a
generalization of a speckle noise model for the Hermitian product of two SAR images $S_{p} S^{*}_{q}$, where $\left\{p, q\right\} \in \left\{hh,hv,vv\right\}$.

This product can be expressed in terms of $\rho = \left|\rho\right|\exp\left(j\phi_x\right), $
the complex correlation coefficient between $S_p$ and $S_q$; $\left|\rho\right|$ is the coherence.

Based on the conclusions of~\cite{Foucher2014}, we used in previous studies~\citep{Palacio17a} the MBPolSAR filter~\citep{Lopez2008}.
This filter considers all the information provided by the covariance matrix, so it improves the estimation of the individual entries and the polarimetric information.
Different processes are required to filter diagonal and off-diagonal elements of $\mathbf{Z}$. 
Multilook filtering is applied to the diagonal elements, while
off-diagonal elements are processed differently according to $\hat{\rho}$, which is calculated as in~\cite{Lopez2007}, in order to reduce the coherence bias. 

In all methods of Table~\ref{table1} we applied the filter with squared windows of sides $3,5,7,\ldots,19$. 
In each situation, we obtained nine channels, that are used as input for the classifier.

\subsection{Classification}
\label{sec:Clasif}

The classification of images gives more information about the scene in the image than just digital values. 

PolSAR classification plays an active role in many domains as a significant part of remote sensing image processing. 
Although many studies have reported various methods with greater classification accuracy using PolSAR data instead of conventional single polarization SAR data,
the way to find an effective classifier is very important for PolSAR image classification~\citep{Aghababaee2013}.

In supervised classification the user supervises the pixel classification process, by selecting representative sample sites of known cover type called Training Areas. 

\cite{Narayanan02} performed the supervised classification using the Maximum Likelihood Classifier (ML)~\citep{Strahler1980}.
ML is a widely used method for classifying remotely sensed data~\citep{Maselli1992}.  
It is based on the Bayes Decision Rule where a pixel is assigned to that class which has a posteriori probability greater than that for all other classes.

Support Vector Machine (SVM) is a supervised classification method well suited to handle linearly nonseparable cases~\citep{Burges1998}, which is the situation at hand.
Among other advantages, it allows defining feature vectors with numerous and heterogeneous components~\citep{SARAutomaticTargetRecognitionUsingJointLowRankandSparseMultiviewDenoising2018}, not requiring the specification of the probabilistic distribution of the data.

\begin{figure}[hbt]
\centering
\includegraphics[width=\linewidth]{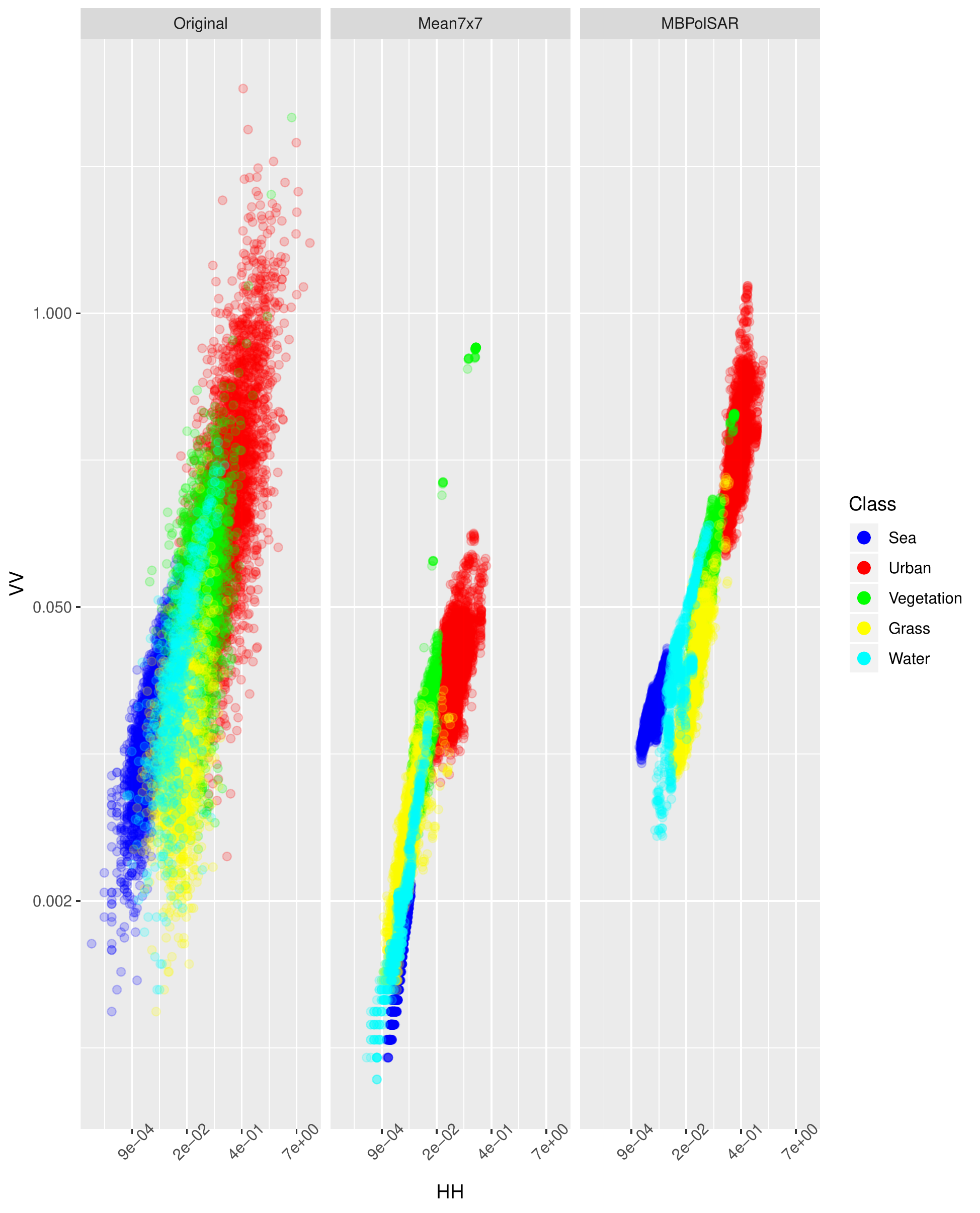}
\caption{Observations from the training samples in the original, mean-filtered and MBPolSAR-filtered images}\label{fig:Samples}
\end{figure}

Fig.~\ref{fig:Samples} shows scatterplots of the data from the training samples in the original image (left, shown in Fig.~\ref{fig:SF}), mean-filtered image (middle), and MBPolSAR-filtered image.
Only the HH (abscissas) and VV (ordinates) components are shown in log-log scale.
The nonlinearity of the classification problem is clear, thus making SVM an appropriate tool for classification.

SVMs were proposed as an automatic learning system based on the statistical learning theory~\citep{vapnik1995nature}. 
This method has captured the attention of researchers due to its successful performance in different areas such as face recognition, textual categorization, predictions, image retrieval and handwriting recognition. 
One of the outstanding features is its excellent generalization ability even in high dimensional spaces and with small training sets.

SVMs have been applied in a number of problems, and their use in remote sensing is spreading, leading to improved results with respect to traditional classifiers like ML~\citep{Aghababaee2013,RemoteSensingImageClassificationaSurveyofSupportVectorMachineBasedAdvancedTechniques2017}.
\cite{Lardeux2009} presented a brief description of a SVM concluding that it performs much better than ML under the Wishart model when applied to an optimized set of polarimetric indicators.
SVMs transform feature vectors into a larger dimension space, where classes can be linearly separated. 
It is an optimization algorithm to determine the optimal boundary between two groups. 
The simplest case is the linearly separable, where there is a positive distance between groups and it is possible to choose a separation hyperplane with maximal distance to each.

Consider two classes $C_1$, $C_2$ with labels $-1$ and $+1$. 

Given $((\textbf{y}_{1\bullet},r_1),(\textbf{y}_{2\bullet},r_2),\ldots,(\textbf{y}_{n\bullet},r_n))$ a training sample of size $n$,  where $r_i=+1$ if $\textbf{y}_{i\bullet}\in C_1$ and $r_i=-1$ otherwise,
with $\textbf{y}_{i\bullet} = (y_{i1}, y_{i2},\ldots, y_{ip})^T \in \mathbb{R}^p$, where $p$ is the size of the feature space. 
The classification problem consists in finding an optimal hyperplane for separating the two classes, which is 
\begin{equation}
g(\textbf{y})=\textbf{w}^T \textbf{y}+ w_0,
\label{hiper}
\end{equation}
where $\textbf{w}$, the support vector, is the perpendicular to the hyperplane and $w_0$  is a scalar. 
The distance from the hyperplane to the closest points of each side is called the \textit{margin} $M$. 
The optimal separation is the one that maximizes the distance $M$, which is calculated as: $ M=( 2\left\|w\right\|)^{-1}$. 
The class to which an unknown observation belongs to is determined by the application of~\eqref{hiper}: 
if the result is positive the observation is assigned to $C_1$, otherwise to $C_2$. 
Strategies have been proposed to solve classification problems with more than two classes (multiclass): one-against-all and one-against-one, here we use the latter.

However, the presence of speckle in SAR images makes classification a non-linearly separable problem; cf.~\citet{Aghababaee2013,RegionBasedClassificationPolSAR_RBF_StochasticDistances}, and Fig.~\ref{fig:Samples}.
In this case the data are mapped onto an $m$-dimensional ($m>p$) space through a transformation $\Phi:\mathbb R^p\rightarrow\mathbb R^m$ to go back to a linear problem.
The analytical form of $\Phi$ is not required, only its scalar product $\Phi\left(\textbf{y}_i\right)^T\Phi\left(\textbf{y}\right)$. This product, called a \textit{kernel} function $K$, is the non-linear projection of the data onto the original space (dimension $p$). 
Then, the corresponding decision function can be expressed in terms of $K$.
There are several kernel functions; in this work we use Radial Basis Functions (RBF) with parameter $\gamma = p^{-1}$, and $p=9$ is the dimension of our feature space.
SVMs also need a penalty value for missclasification errors, usually called $C$; in this work we used $C=1$.

Some authors consider a fixed parameter space and search for the configuration of $(C,\gamma)$ which produces the most accurate results with respect to testing samples.
For simplicity and reproducibility, we used the default values of the \verb|libsvm| library in the \verb|e1071| package of the \texttt R platform~\citep{ManuR}.
	
For all methods of Table~\ref{table1} we identified the same locations of training areas which, in turn, are different from the testing areas that are used to the classification performance stage.

\subsection{Classification with filtered image as input}\label{sec:Aplic1}

The procedure of classification used for each method of Table~\ref{table1} consists of:
\begin{enumerate}
	\item Applying the filter (LM or MBPolSAR) with sliding windows.
	\item Selecting as features for classification the nine channels resulting from the filtering.
	\item Extracting training samples from the filtered image. 
	\item Predicting all pixels of the image by the classifier (ML or SVM) and obtaining the thematic map.
\end{enumerate}
The procedure is repeated for each window size.

\subsection{Validation}
\label{sec:Valid}

To evaluate the performance of the proposed methods quantitatively we constructed a Confusion Matrix for each situation (windows of sizes $3,5,7,\ldots,19$) and each method, using the same testing data set.
From these matrices we calculate measures of map precision:
\begin{enumerate}[label=({\alph*})]
	\item Overall Accuracy coefficient~(F), as the proportion of pixels correctly classified;
	\item Kappa coefficient ($\kappa$), as the proportion of pixels correctly classified considering all the matrix entries;
	\item Classification Accuracy coefficient~(CA), as the proportion of pixels correctly classified for each class.
\end{enumerate}
We also evaluate qualitatively the thematic maps by visual inspection.

The computing platform was a combination of Open Source and licensed software. 
The open source PolSARPro v5.1 software~\citep{ManuPol} was used for the pre-processing stage.
ENVI version 4.8~\citep{MENVI} was the platform of choice for extracting and validating Regions of Interest (ROIs).
All the other analyses were implemented in the R version 3.5.0~\citep{ManuR} platform.

\section{EXPERIMENTAL RESULTS}
\label{sec:Aplic}

We used two images: a \num{550x645} pixels image obtained from the AIRSAR image of San Francisco, and a \num{700x1100} pixels image obtained from a full polarimetric UAVSAR image of Bell Ville city, in C\'ordoba province, Argentina.

\subsection{San Francisco image}

\begin{figure}[hbt]
	\centering
	\includegraphics[width=.6\linewidth,trim ={0cm 3.7cm 0cm 3.8cm}]{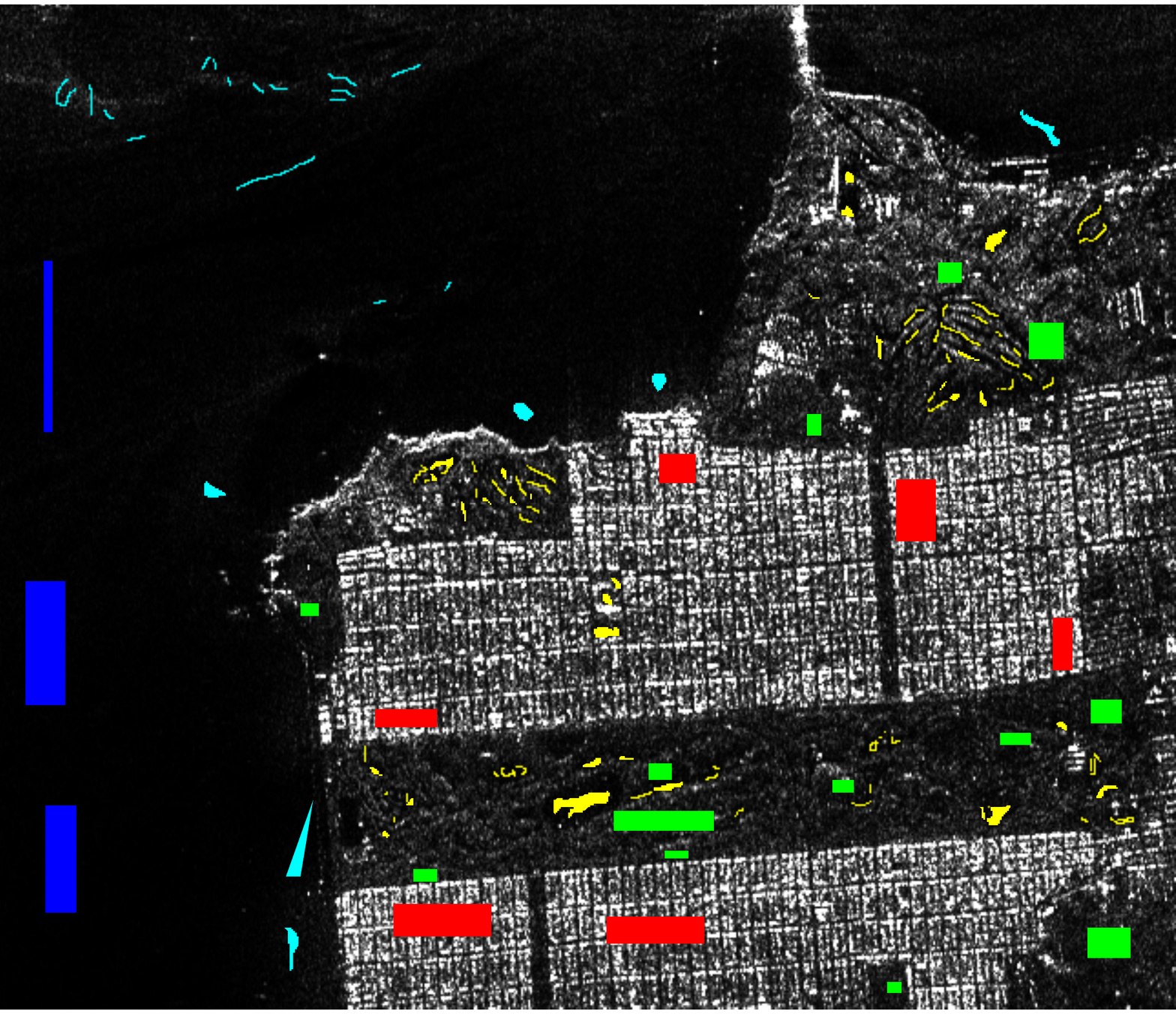}
	\caption{AIRSAR HH data and training samples.\label{fig:SF}}
\end{figure}

AIRSAR was an airborne mission with PolSAR capabilities, designed and built by the Jet Propulsion Laboratory. 
Fig.~\ref{fig:SF} shows a \num{550x645} pixels region obtained from a \num{900x1024} pixels full polarimetric actual image of San Francisco recorded by this sensor in the L~band, acquired with four nominal looks.

We identified $5$ different cover types: Grass, Sea, Urban, Vegetation and Water (areas in the Sea with water movement); their training samples are shown in Yellow, Blue, Red, Green and Cyan, respectively.
Furthermore, for testing classification performance, test samples were also selected.

\begin{figure}[htb]
	\centering
	\subfigure[Method 1 \label{global1}]{\includegraphics[width=.4\linewidth,trim ={0cm 0.1cm 0cm 0.7cm},clip,keepaspectratio]{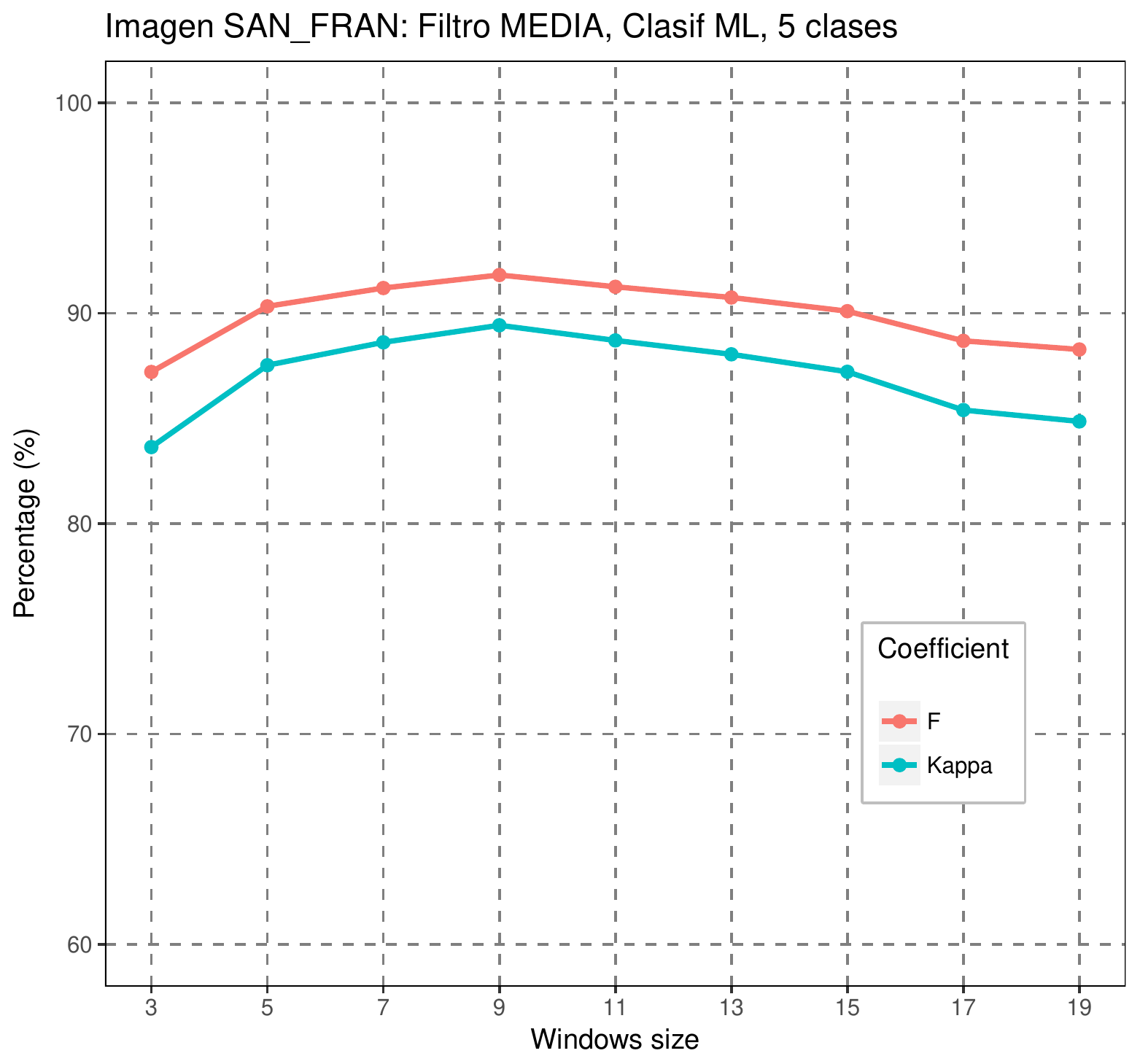}}
	\subfigure[Method 2 \label{global2}]{\includegraphics[width=.4\linewidth,trim ={0cm 0.1cm 0cm 0.7cm},clip,keepaspectratio]{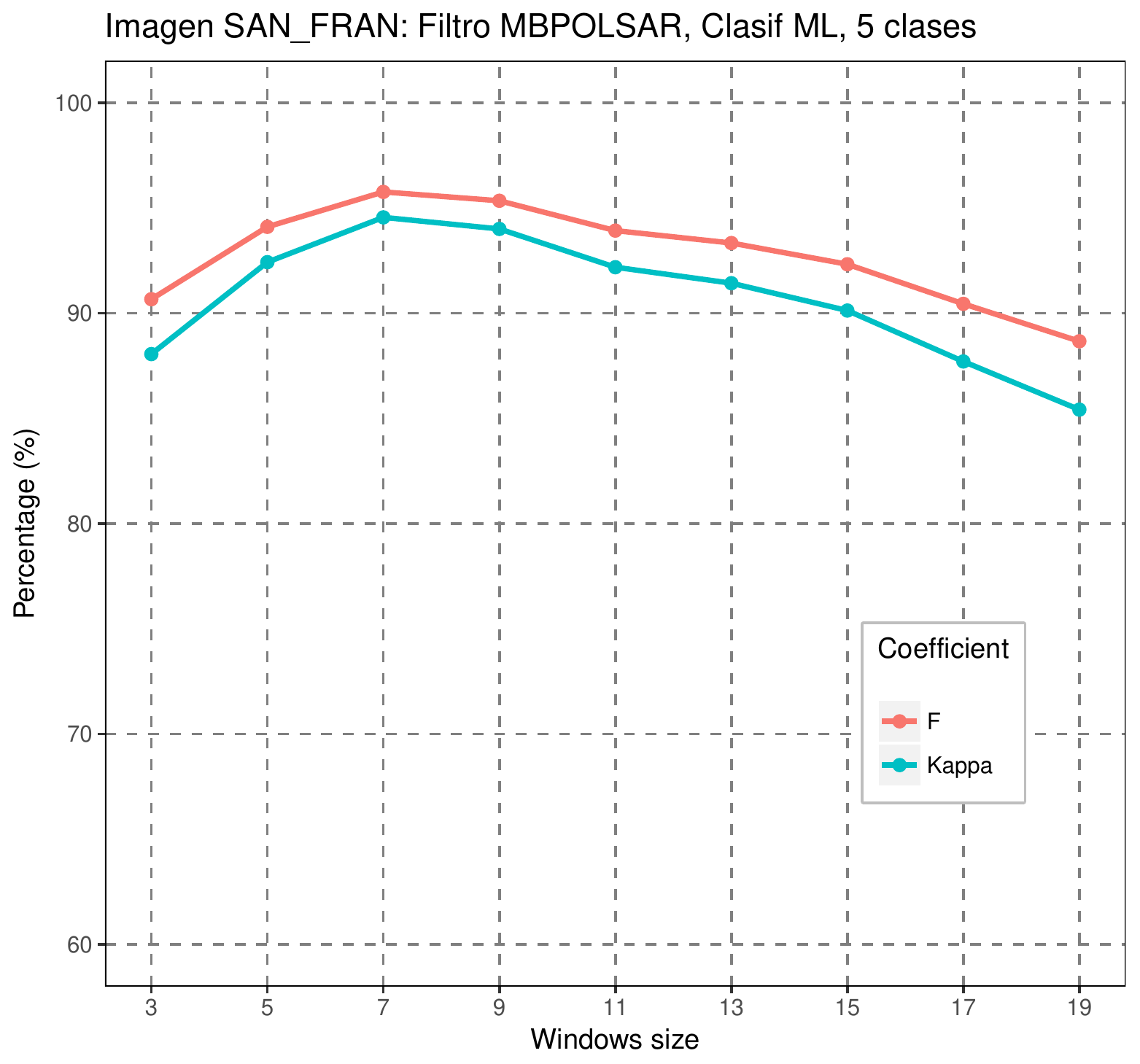}}
	\subfigure[Method 3 \label{global3}]{\includegraphics[width=.4\linewidth,trim ={0cm 0.1cm 0cm 0.7cm},clip,keepaspectratio]{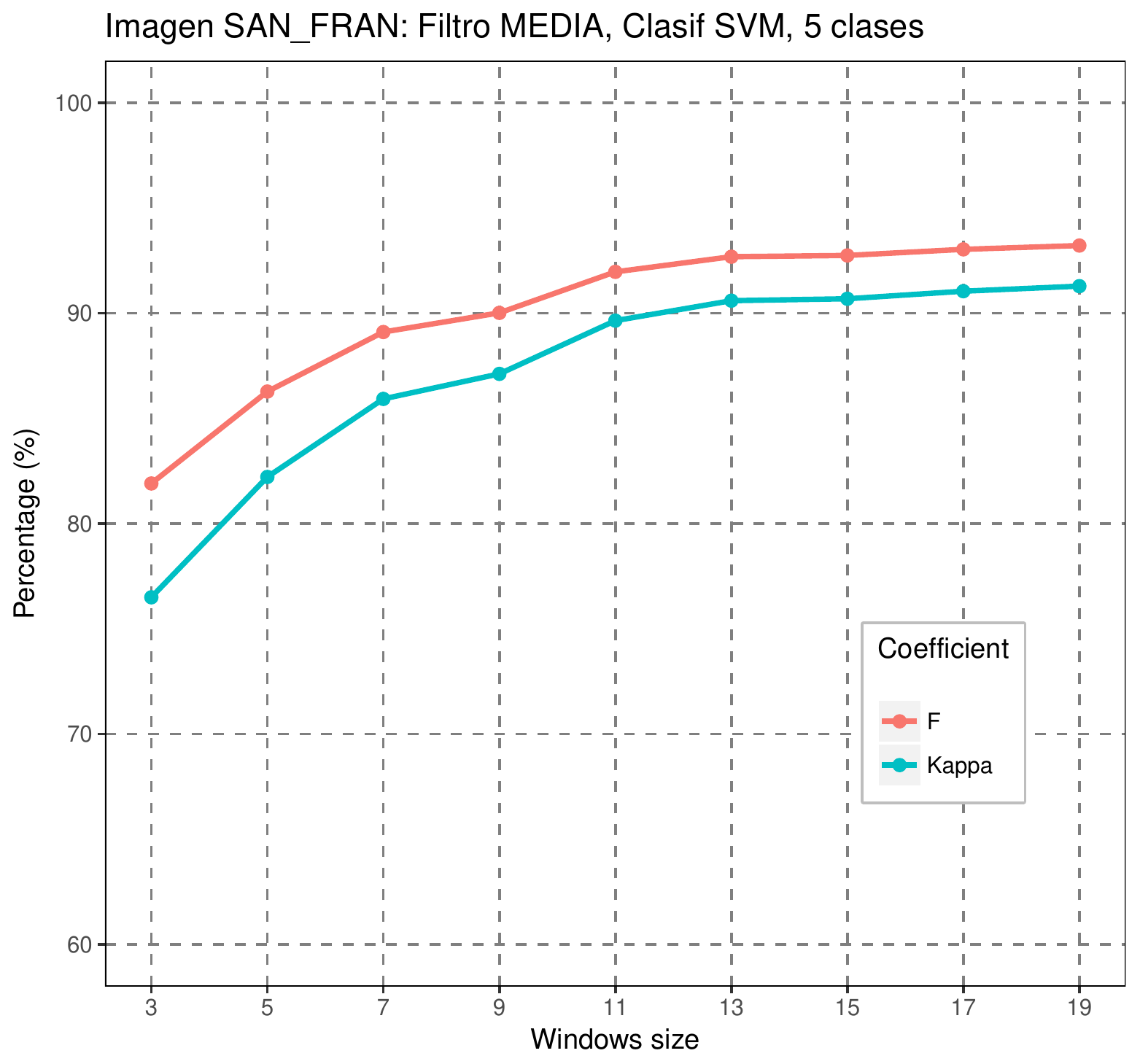}}
	\subfigure[Method 4 \label{global4}]{\includegraphics[width=.4\linewidth,trim ={0cm 0.1cm 0cm 0.7cm},clip,keepaspectratio]{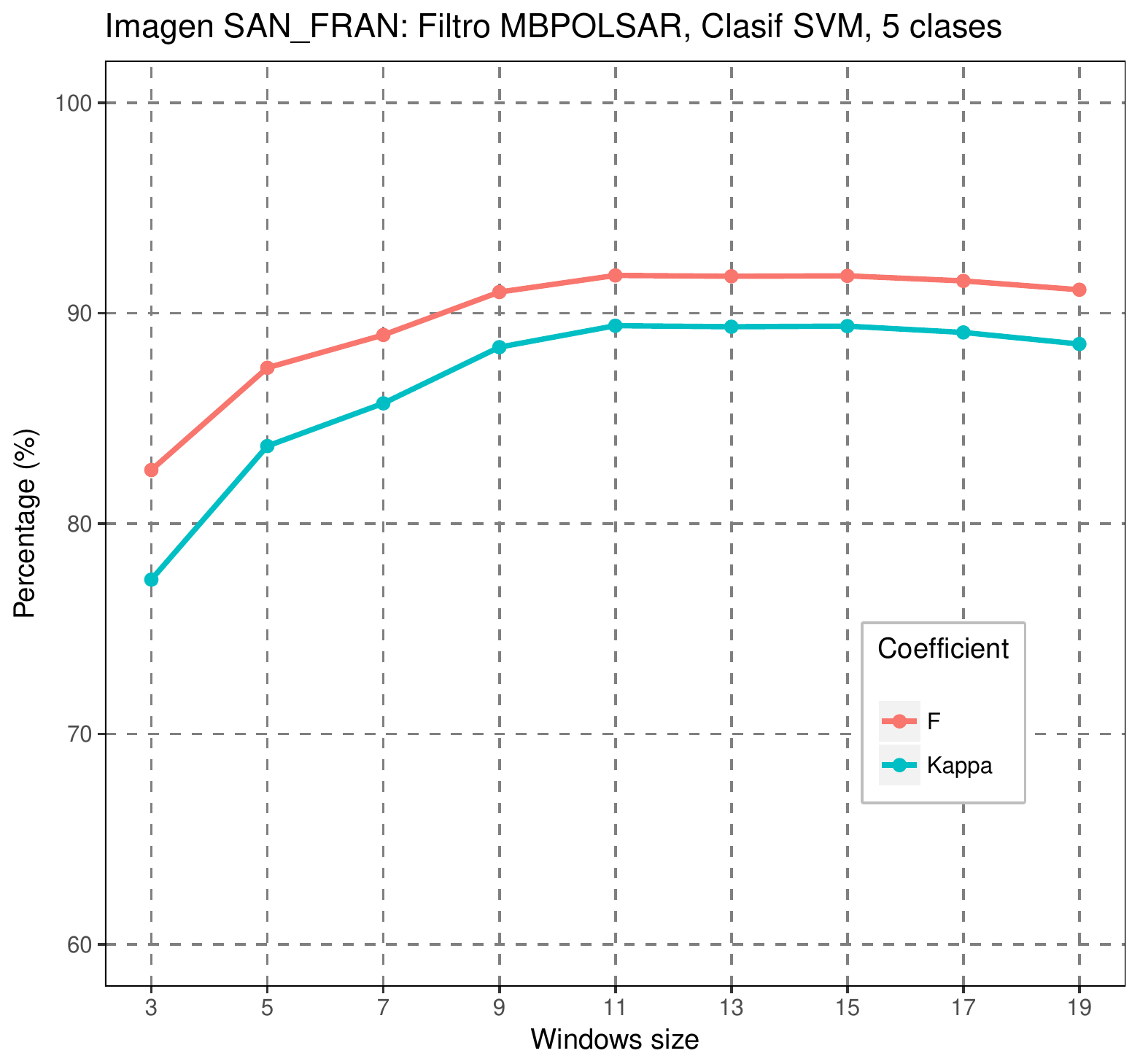}}
	\caption{AIRSAR data. Classification performance by window size.\label{fig:globalSF}}
\end{figure}

Fig.~\ref{fig:globalSF} show that globally the F and Kappa values are similar.
Filtering improves the results but the perfomances are not the same.

Applying ML classifier,
\begin{itemize}
	\item the highest values are obtained with MBPolSAR filter (Fig.~\ref{global2}); 
	\item with Mean filter the maximum value is obtained with window \num{9x9} (Fig.~\ref{global1});
	\item with MBPolSAR filter the maximum value is obtained with window \num{7x7} (Fig.~\ref{global2}).
\end{itemize}

Applying SVM classifier,
\begin{itemize}
	\item  changing window \num{3x3} to \num{5x5} produces the most noticeable difference;
	\item with Mean filter the maximum value is obtained with window \num{19x19} (Fig.~\ref{global3});
	\item with MBPolSAR filter the maximum value is obtained with window \num{11x11} (Fig.~\ref{global4}).
\end{itemize}

Comparing the $4$ methods, 
\begin{itemize} 
	\item the worst situation is with small windows and SVM classifier, however the values increase and reach similar performance of Method $1$ with window \num{9x9};
	\item the best results are obtained with MBPolSAR filter, ML classifier and windows from \num{5x5} to \num{9x9}.
\end{itemize}

The most noticeable difference between SVM and ML is that the first has the best performance with greater degradation, whereas such large windows spoil the ML classification.

\begin{figure}[htb]
	\centering
	\subfigure[Method 1 \label{porclase1}]{\includegraphics[width=.4\linewidth,trim ={0cm 0.1cm 0cm 0.7cm},clip,keepaspectratio]{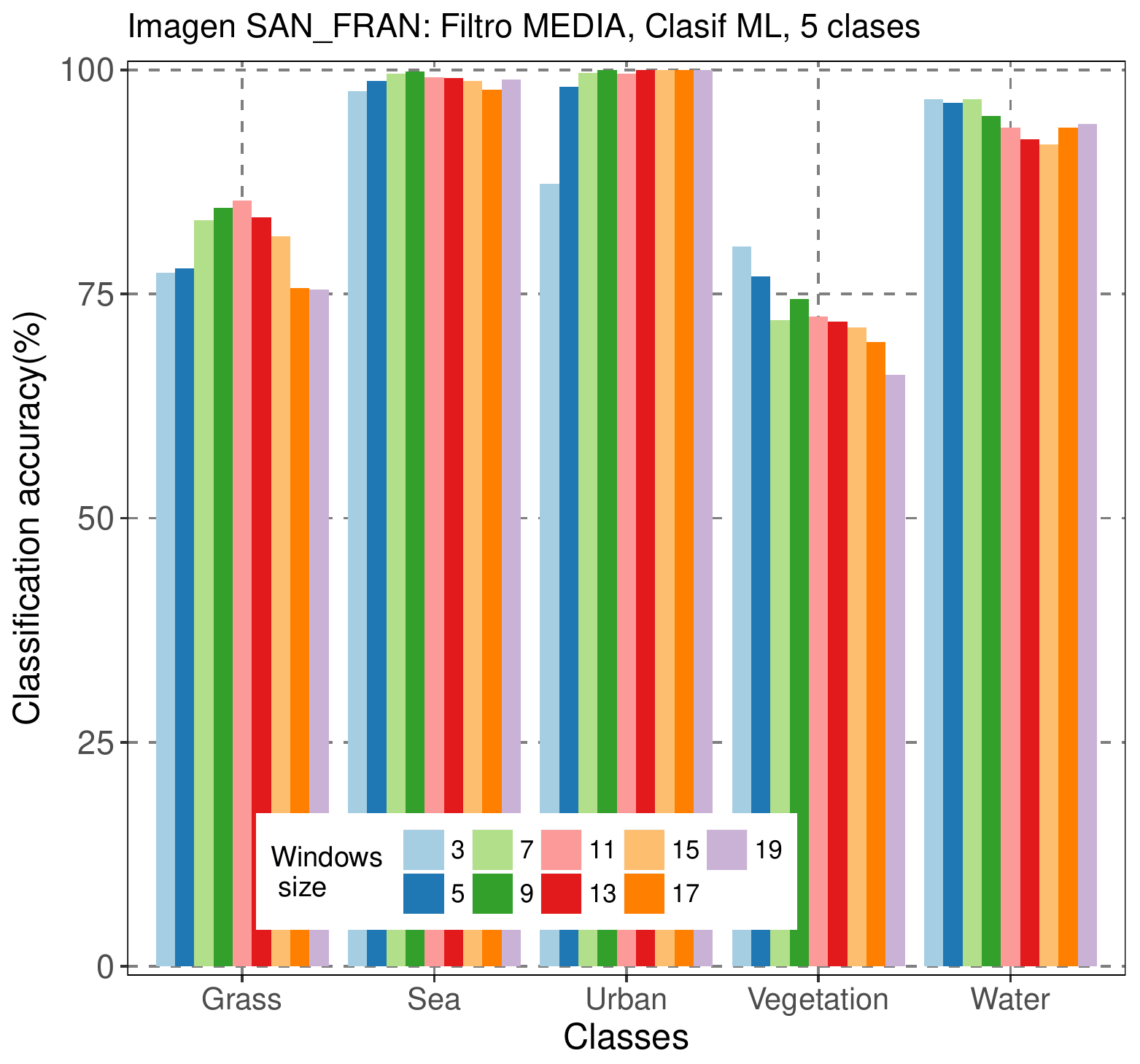}}
	\subfigure[Method 2 \label{porclase2}]{\includegraphics[width=.4\linewidth,trim ={0cm 0.1cm 0cm 0.7cm},clip,keepaspectratio]{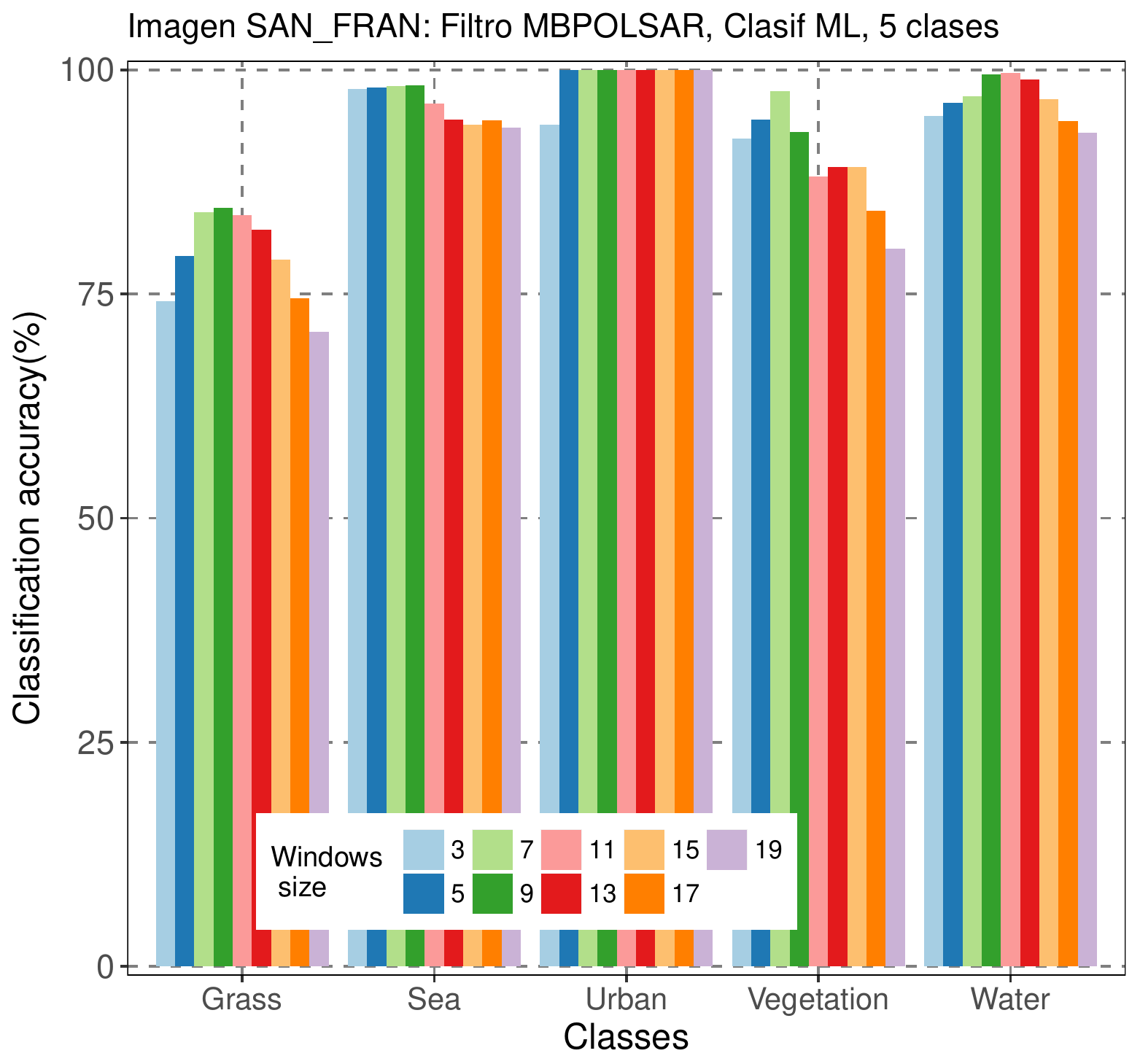}}
	\subfigure[Method 3 \label{porclase3}]{\includegraphics[width=.4\linewidth,trim ={0cm 0.1cm 0cm 0.7cm},clip,keepaspectratio]{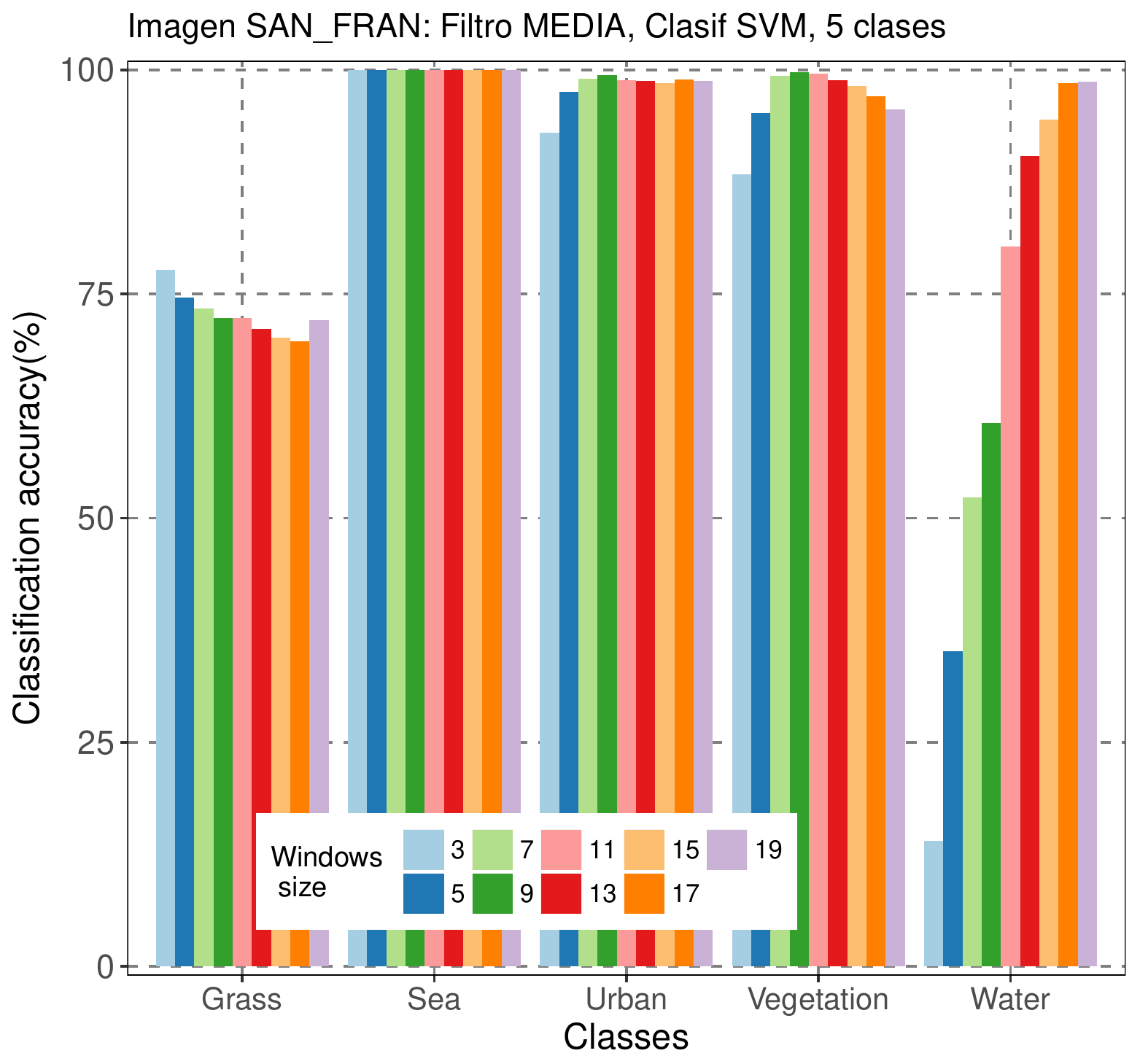}}
	\subfigure[Method 4 \label{porclase4}]{\includegraphics[width=.4\linewidth,trim ={0cm 0.1cm 0cm 0.7cm},clip,keepaspectratio]{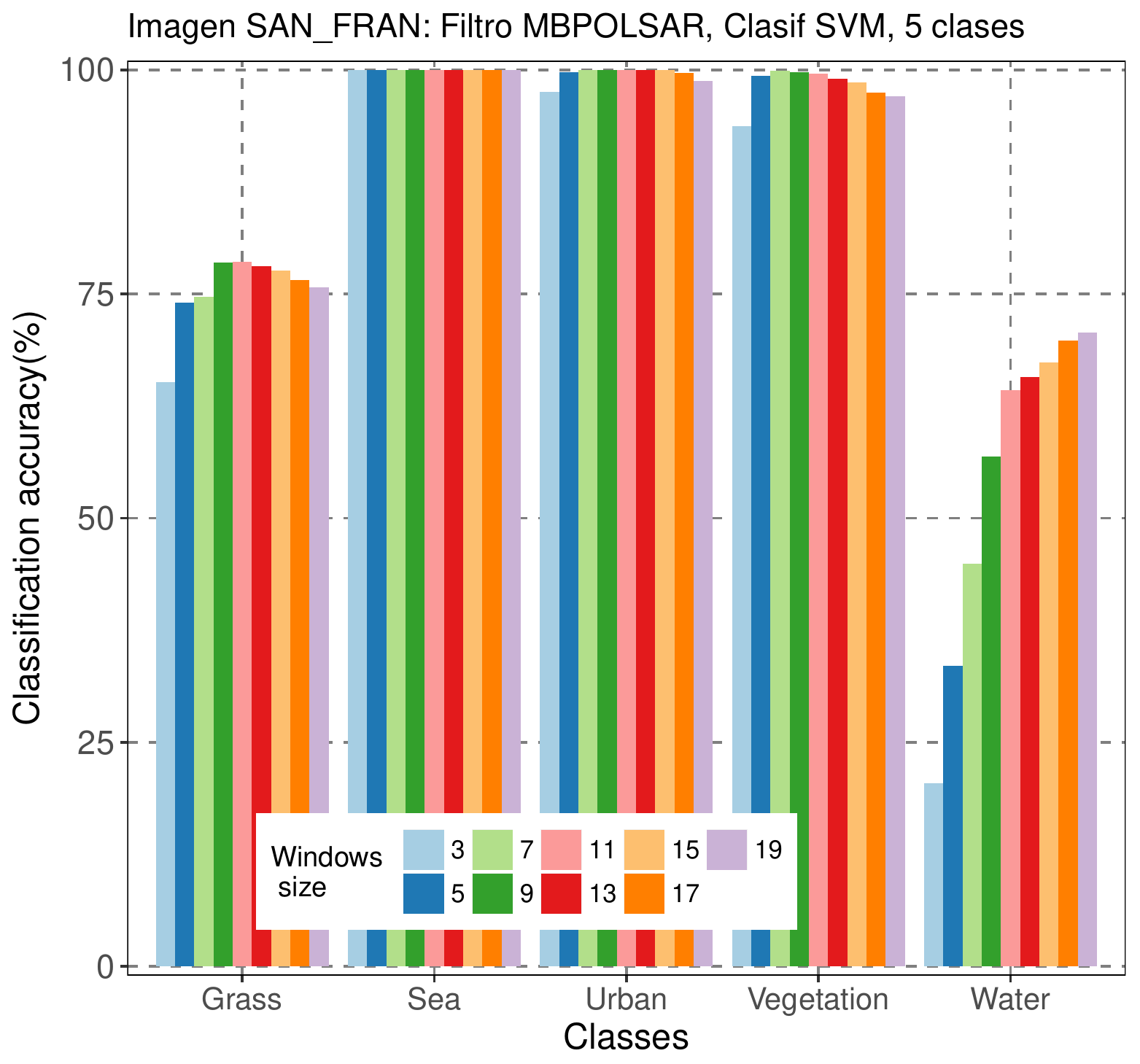}}
	\caption{AIRSAR data. Classification performance by class.\label{fig:porclaseSF}}
\end{figure}

Fig.~\ref{fig:porclaseSF} shows the analysis of classes individually.
The performances are similar when the same classifier is applied.
Classification accuracy takes the maximum values for Sea and Urban classes, except with \num{3x3} window and ML classifier;
in this situation the Water is the best classified.

Numerically, Water class is well classified with ML, whereas with SVM it is the worst classified (with MBPolSAR the values are less than  \SI{75}{\percent}).

\begin{figure}[htb]
	\centering
	\subfigure[Method 1 \label{clasif1}]{\includegraphics[width=.4\linewidth,trim ={0cm 2cm 0cm 2cm},clip,keepaspectratio]{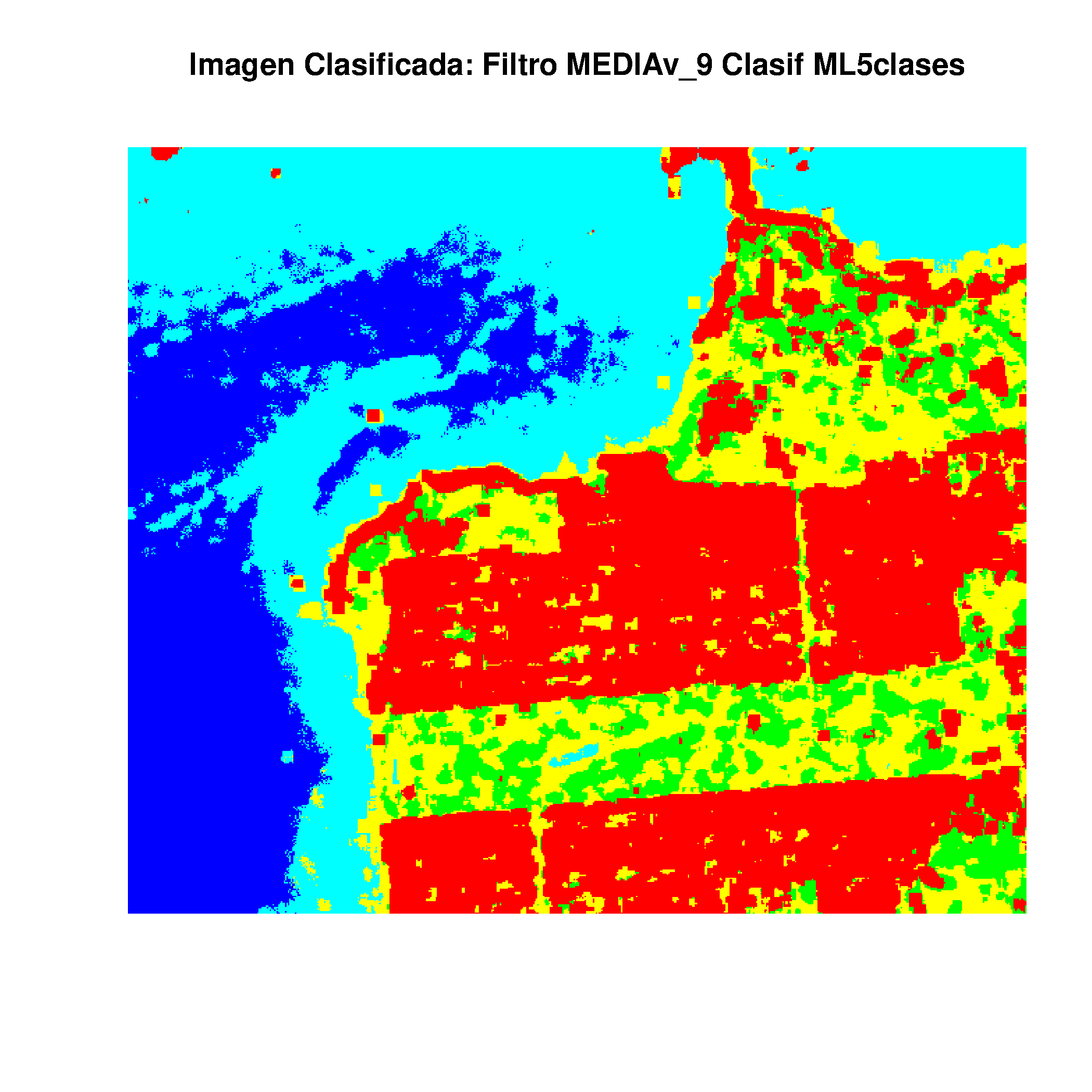}}
	\subfigure[Method 2 \label{clasif2}]{\includegraphics[width=.4\linewidth,trim ={0cm 2cm 0cm 2cm},clip,keepaspectratio]{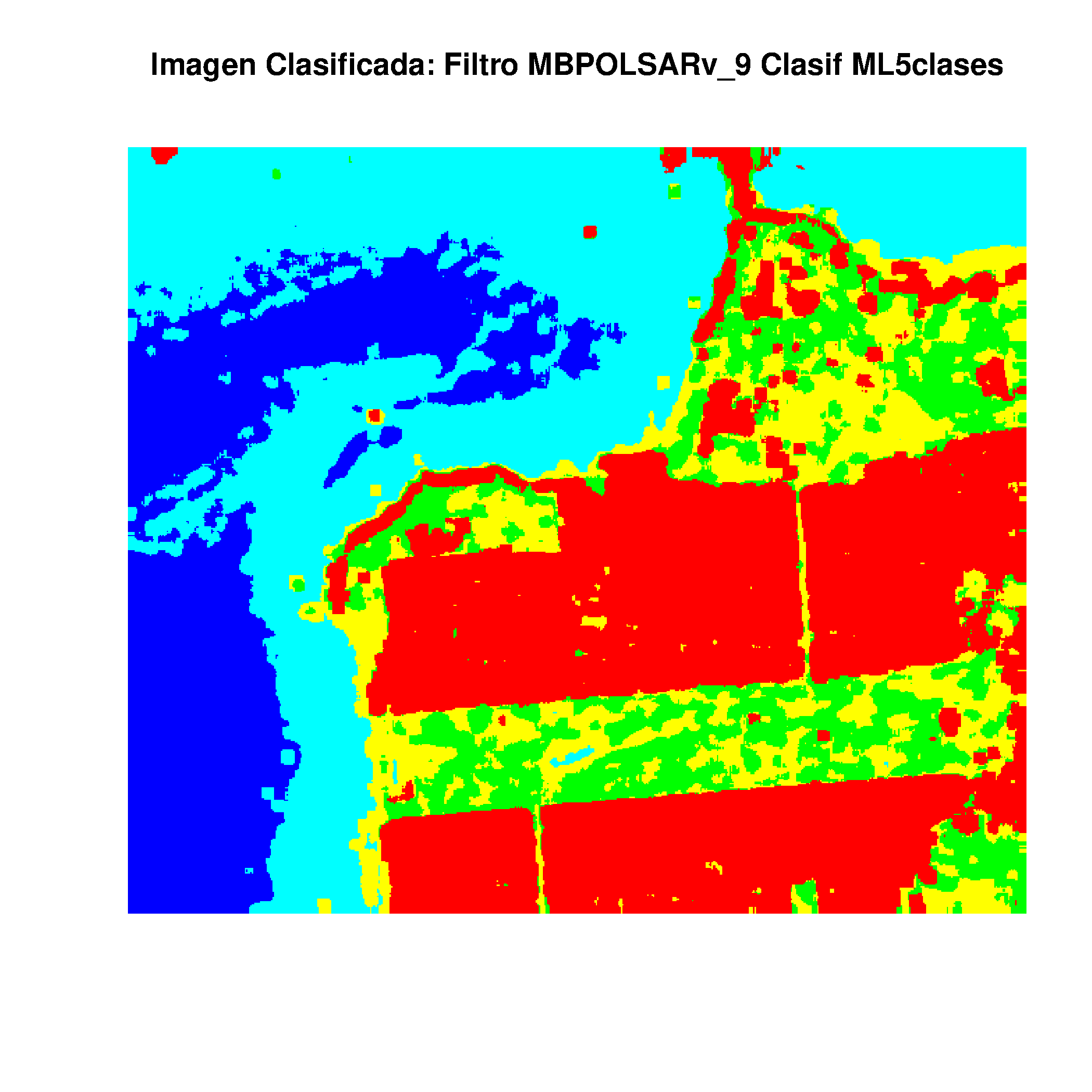}}
	\subfigure[Method 3 \label{clasif3}]{\includegraphics[width=.4\linewidth,trim ={0cm 2cm 0cm 2cm},clip,keepaspectratio]{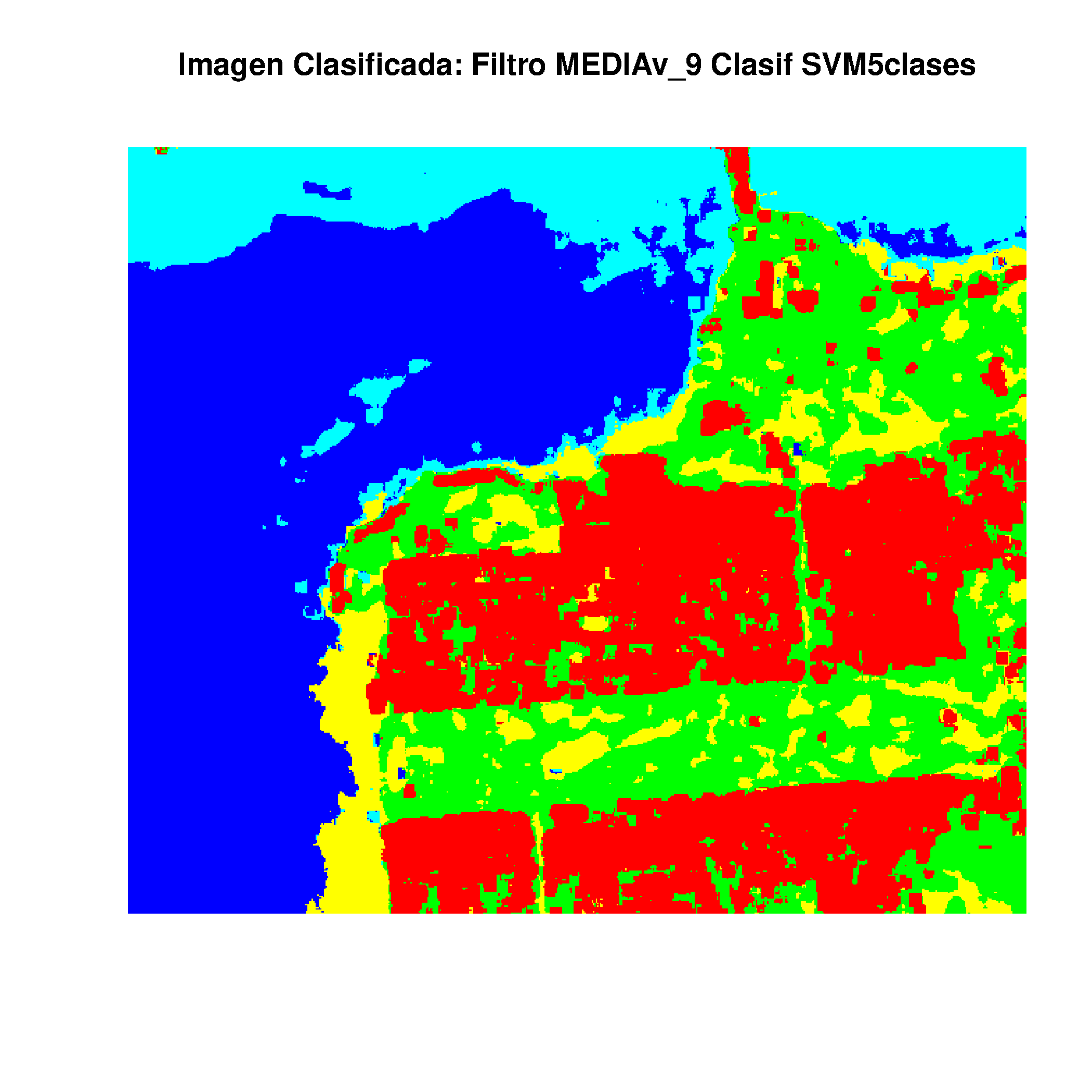}}
	\subfigure[Method 4 \label{clasif4}]{\includegraphics[width=.4\linewidth,trim ={0cm 2cm 0cm 2cm},clip,keepaspectratio]{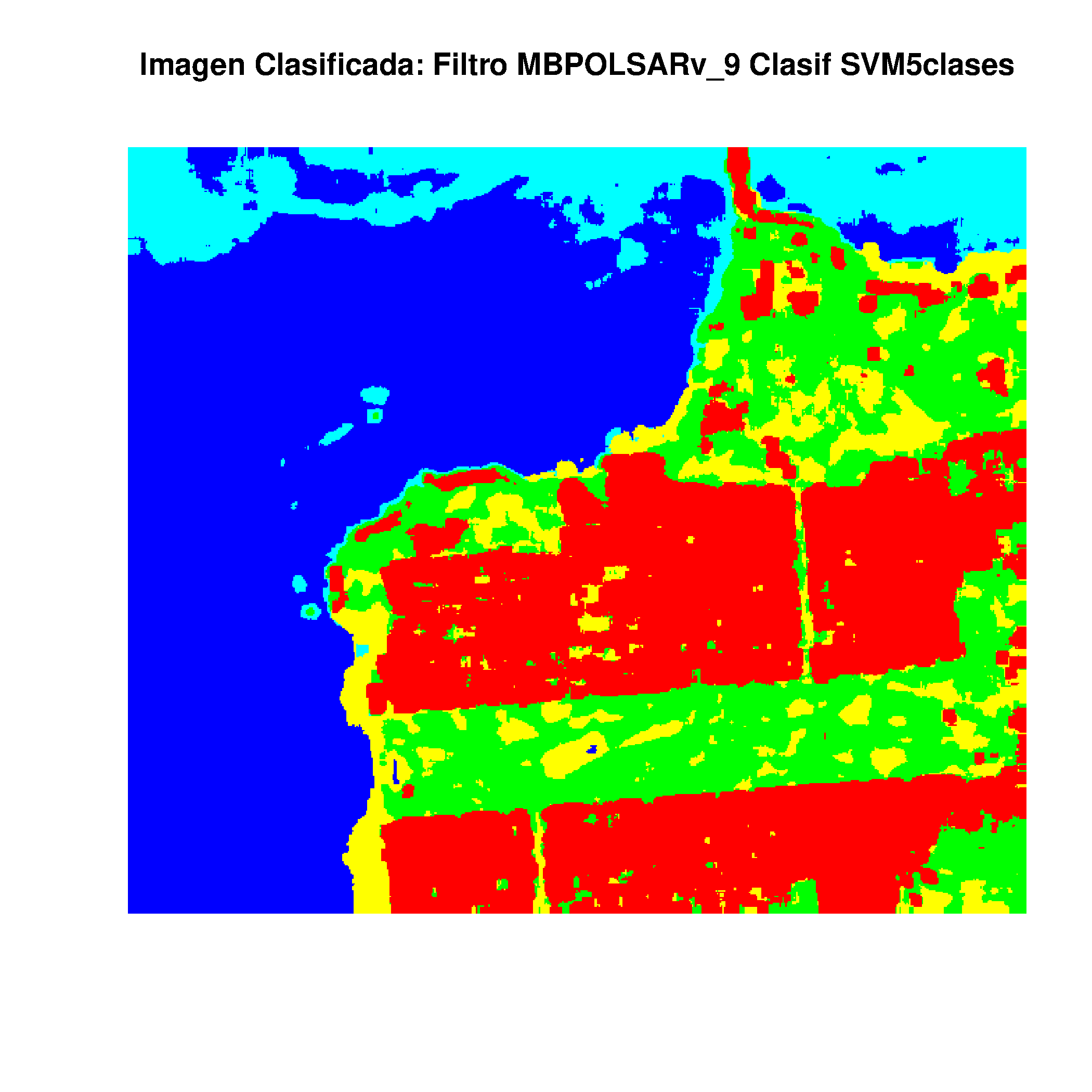}}
	\caption{AIRSAR data. Thematic maps using \num{9x9} windows.\label{fig:mapasSF}}
\end{figure}

However, due to the texture of the sea, almost all the maps produced by ML wrongly allocate to Water pixels of Sea class, Fig.~\ref{fig:mapasSF}.
Furthermore this classifier leads to great loss of detail in the Grass class.

\subsection{Bell Ville image}

Finally, we apply the methods to the UAVSAR image of C\'ordoba province, Argentina, recorded in the L-band, acquired with three nominal looks.
We identified four different cover types:  Urban, Water, Trees and Culture; their training samples are shown in Red, Blue, Green and Orange.

\begin{figure}[hbt]
	\centering
	\includegraphics[width=.7\linewidth,trim ={0cm 6.7cm 0cm 3.3cm}]{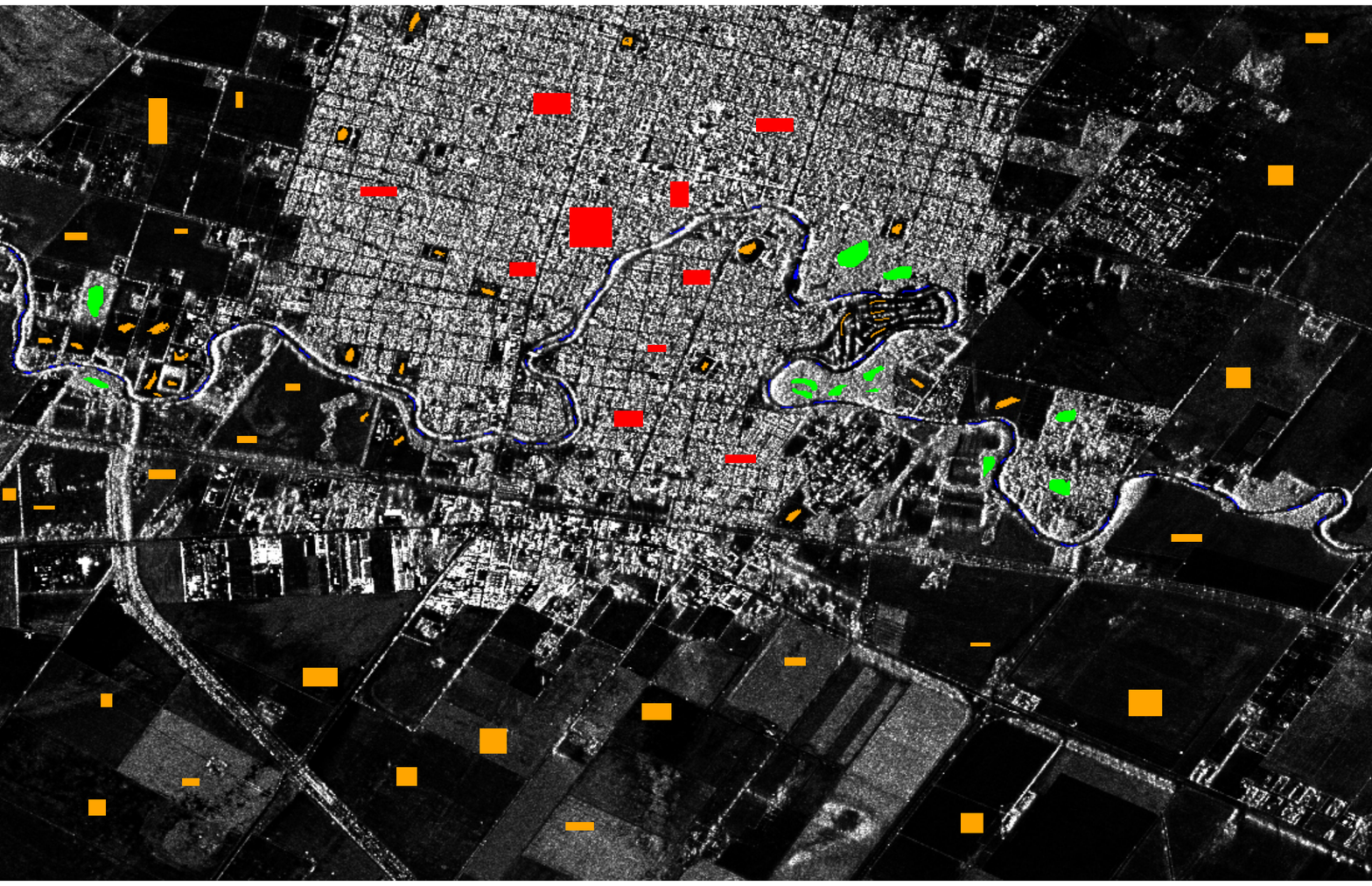}
	\caption{UAVSAR data and training samples.\label{fig:bell}}
\end{figure}

Fig.~\ref{fig:bell} shows the data with training samples.

\begin{figure}[htb]
	\centering
	\subfigure[Method 1 \label{global5}]{\includegraphics[width=.4\linewidth,trim ={0cm 0.1cm 0cm 0.7cm},clip,keepaspectratio]{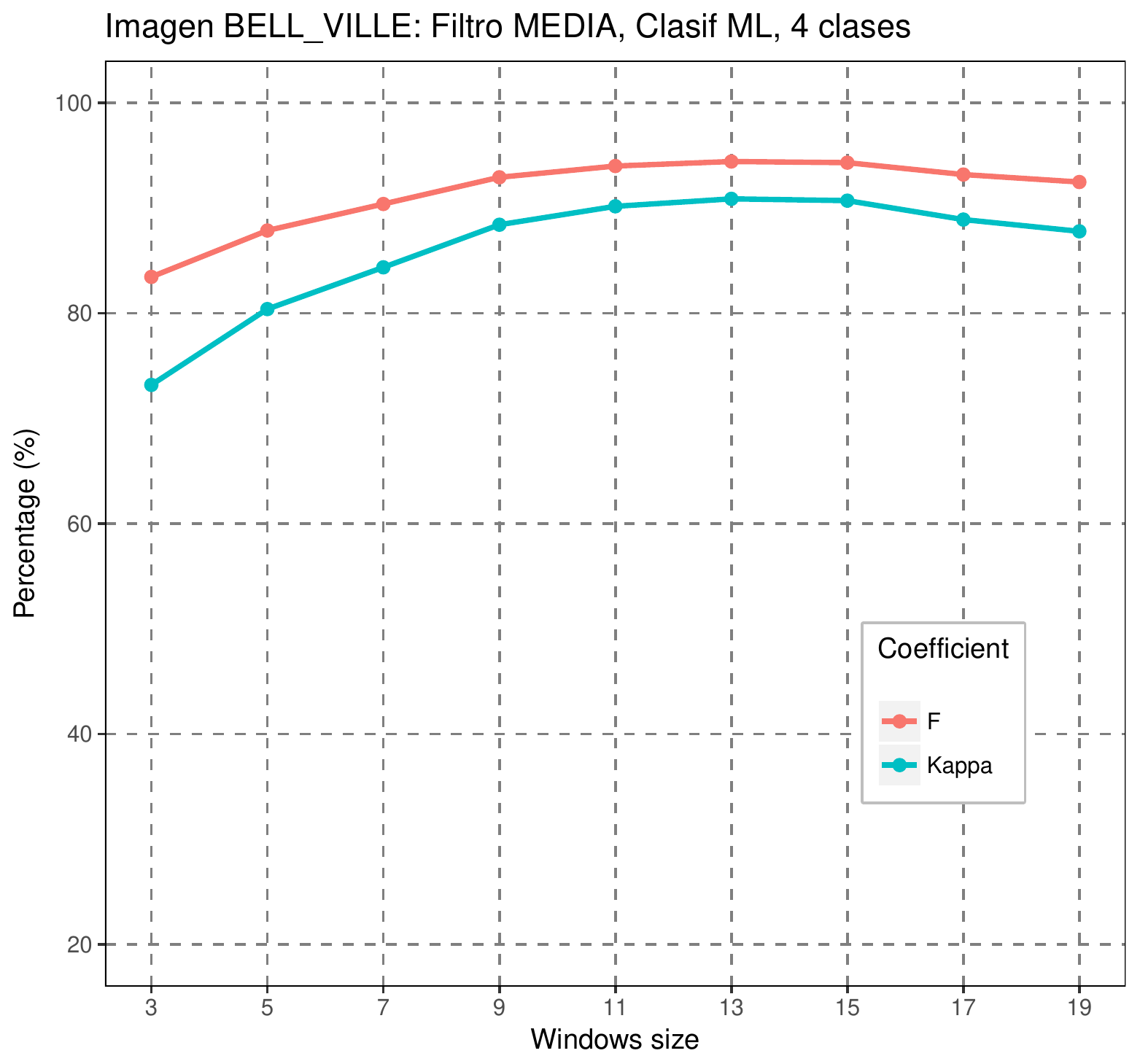}}
	\subfigure[Method 2 \label{global6}]{\includegraphics[width=.4\linewidth,trim ={0cm 0.1cm 0cm 0.7cm},clip,keepaspectratio]{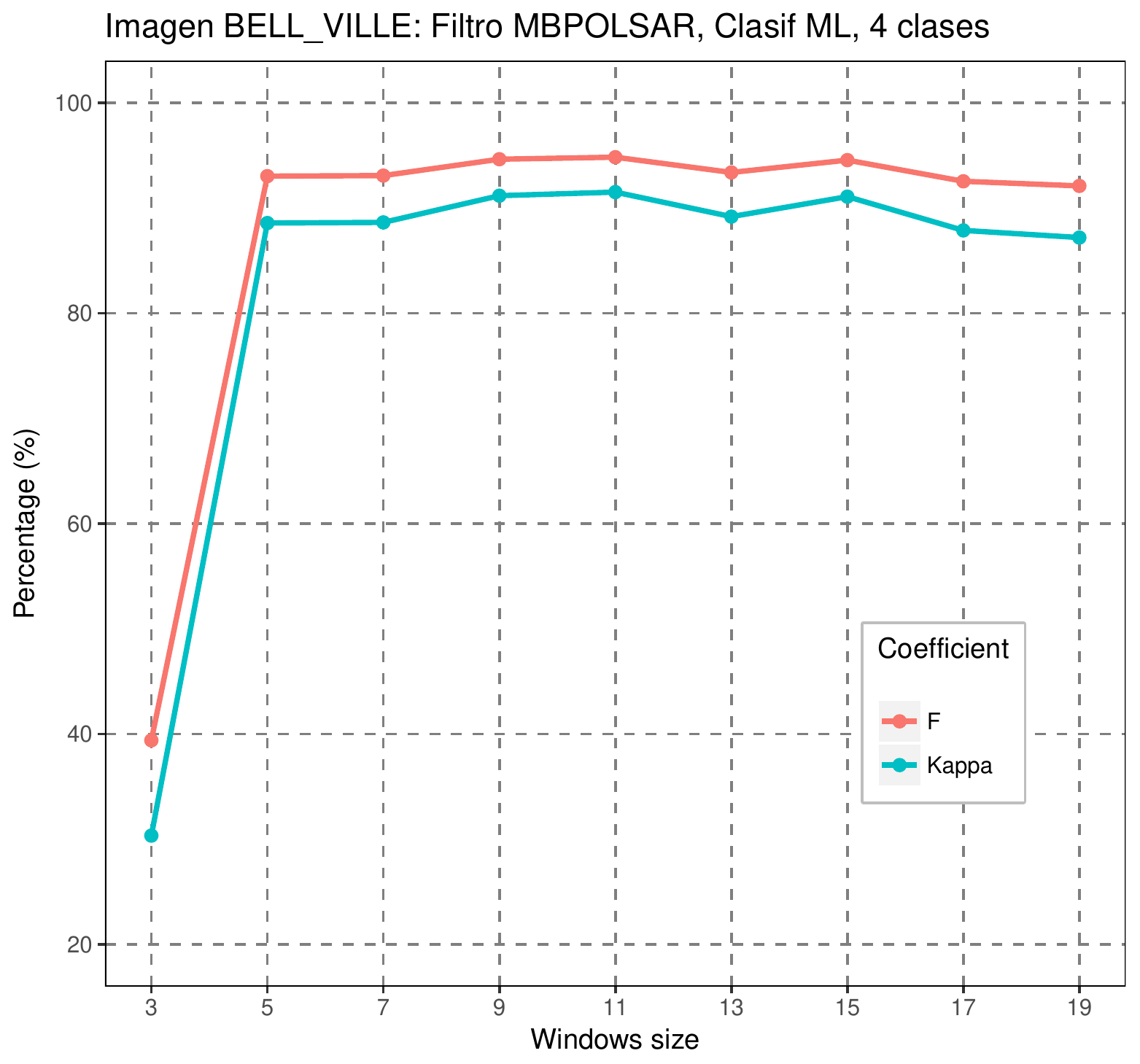}}
	\subfigure[Method 3 \label{global7}]{\includegraphics[width=.4\linewidth,trim ={0cm 0.1cm 0cm 0.7cm},clip,keepaspectratio]{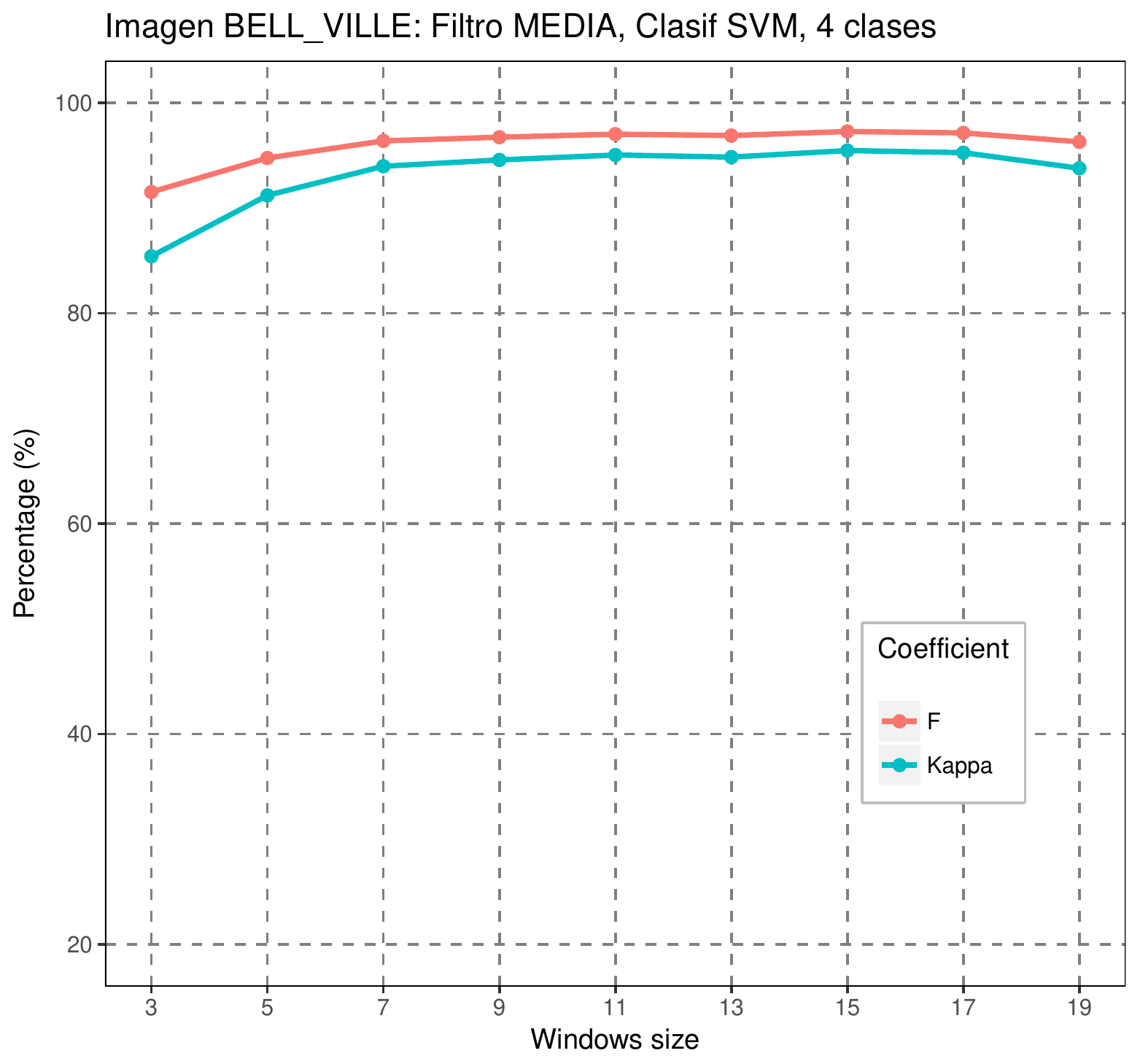}}
	\subfigure[Method 4 \label{global8}]{\includegraphics[width=.4\linewidth,trim ={0cm 0.1cm 0cm 0.7cm},clip,keepaspectratio]{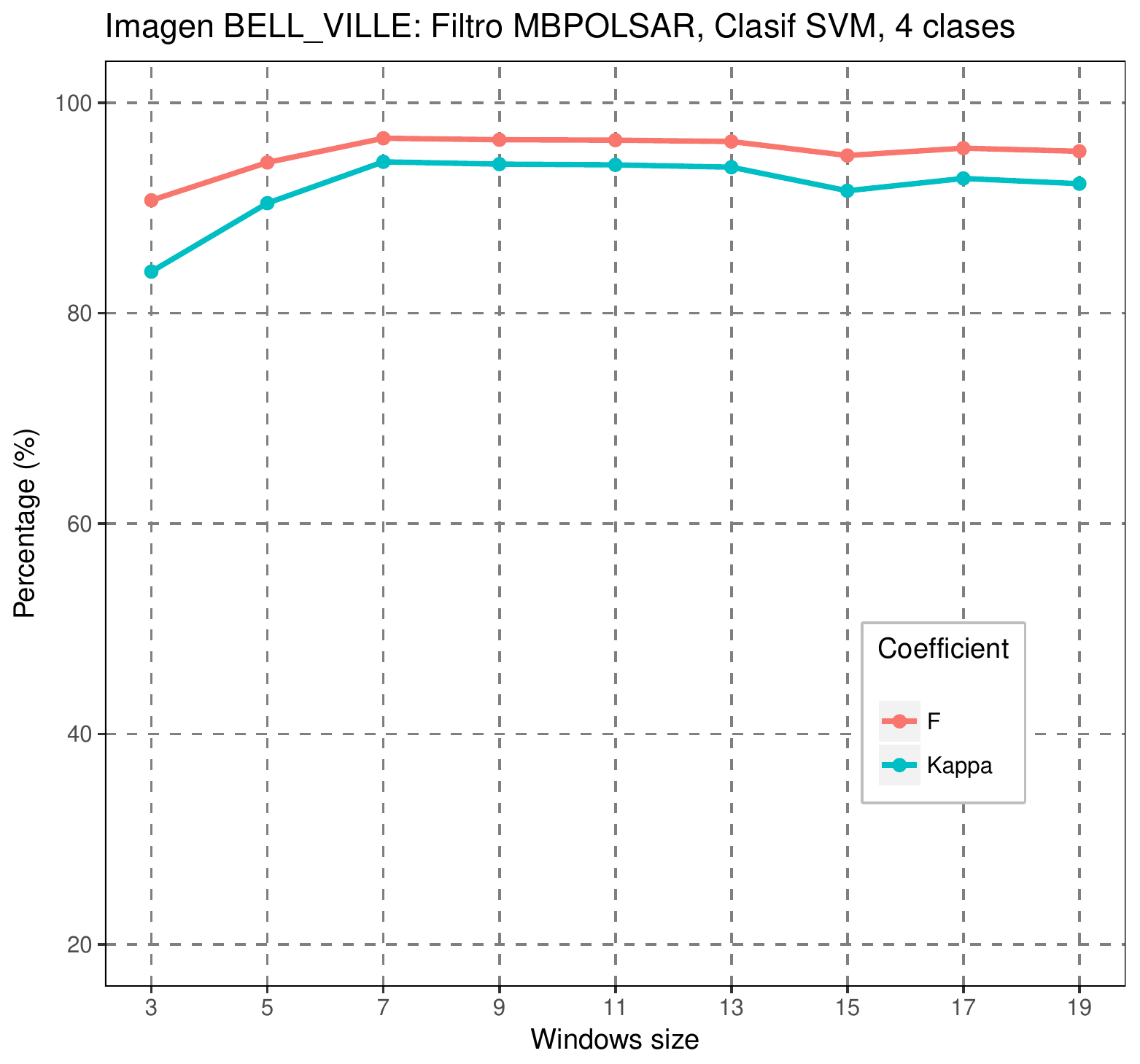}}
	\caption{UAVSAR data. Classification performance by window size.\label{fig:globalBELL}}
\end{figure}

Globally (Fig.~\ref{fig:globalBELL}) the F and Kappa values are similar.
Filtering improves the results, specially changing the window size from \num{3x3} to \num{5x5}.

The worst situation is with Method $2$  (Fig.~\ref{global6}) and window \num{3x3}, however the values increase and reach similar performance of the other methods.
Almost all values are high (except for Method $2$ window \num{3x3}) and increase with degradation.

Nevertheless, with Mean filter (Figs.~\ref{global5} and~\ref{global7}) the increase is most noticeable initially, up to \num{7x7} or \num{9x9} windows, and are slightly reduced for larger windows.
Whereas with MBPolSAR filter (Figs.~\ref{global6} and~\ref{global8}) the performance is unstable, especially when it is classified with ML, although the variations are not very remarkable.

With windows larger than \num{9x9} all the methods show good and similar results.

\begin{figure}[htb]
	\centering
	\subfigure[Method 1 \label{porclase5}]{\includegraphics[width=.4\linewidth,trim ={0cm 0.1cm 0cm 0.7cm},clip,keepaspectratio]{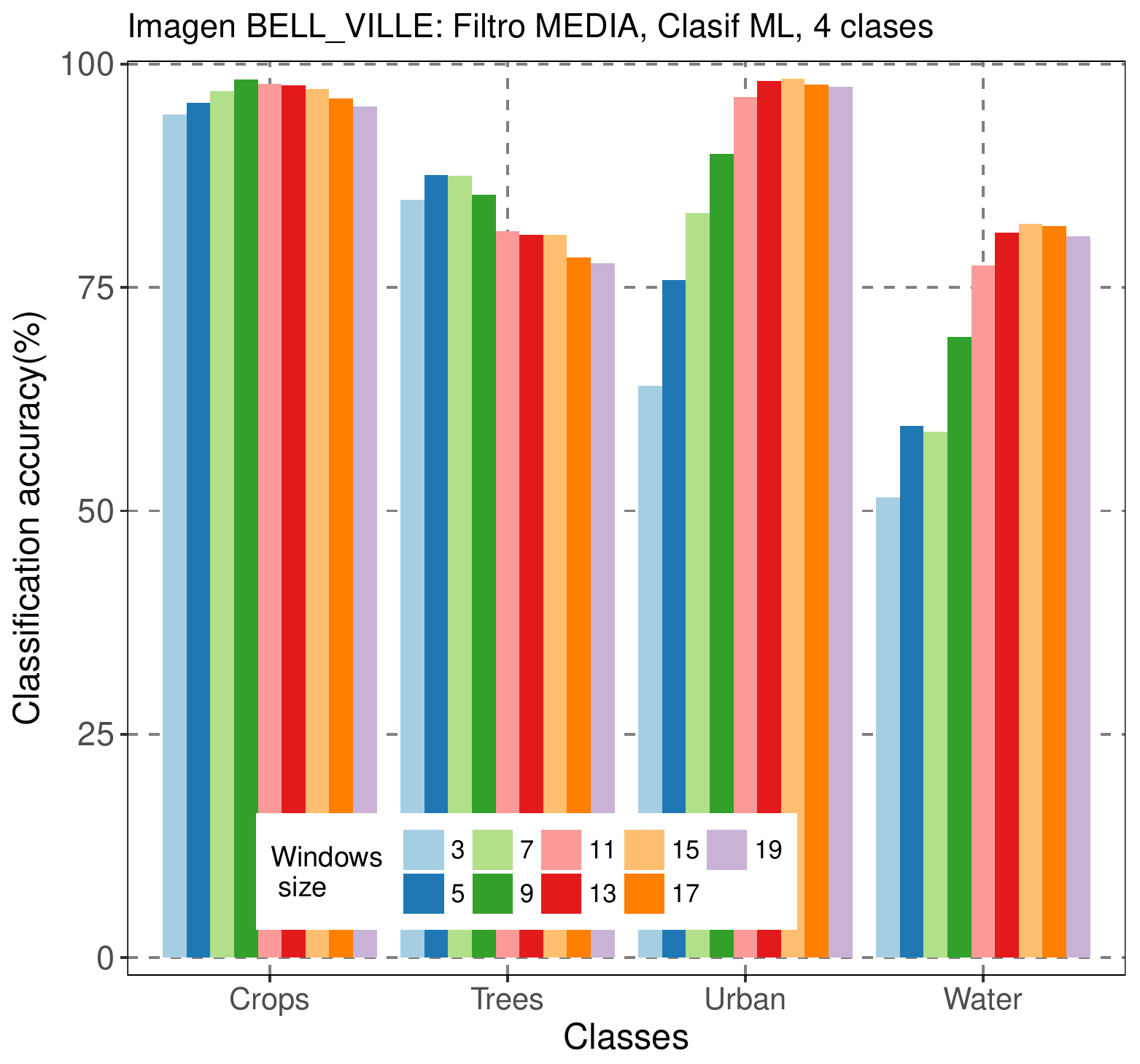}}
	\subfigure[Method 2 \label{porclase6}]{\includegraphics[width=.4\linewidth,trim ={0cm 0.1cm 0cm 0.7cm},clip,keepaspectratio]{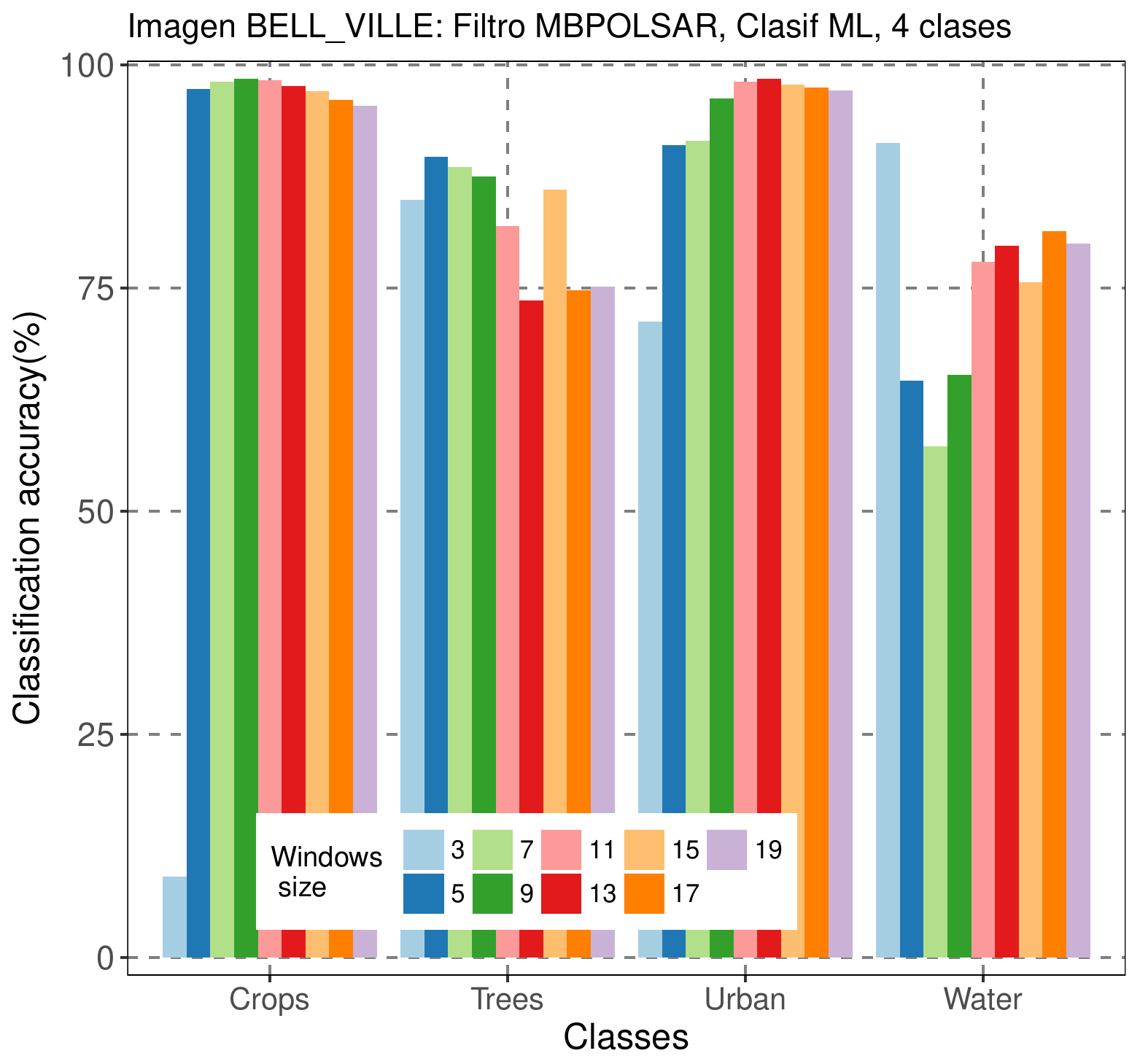}}
	\subfigure[Method 3 \label{porclase7}]{\includegraphics[width=.4\linewidth,trim ={0cm 0.1cm 0cm 0.7cm},clip,keepaspectratio]{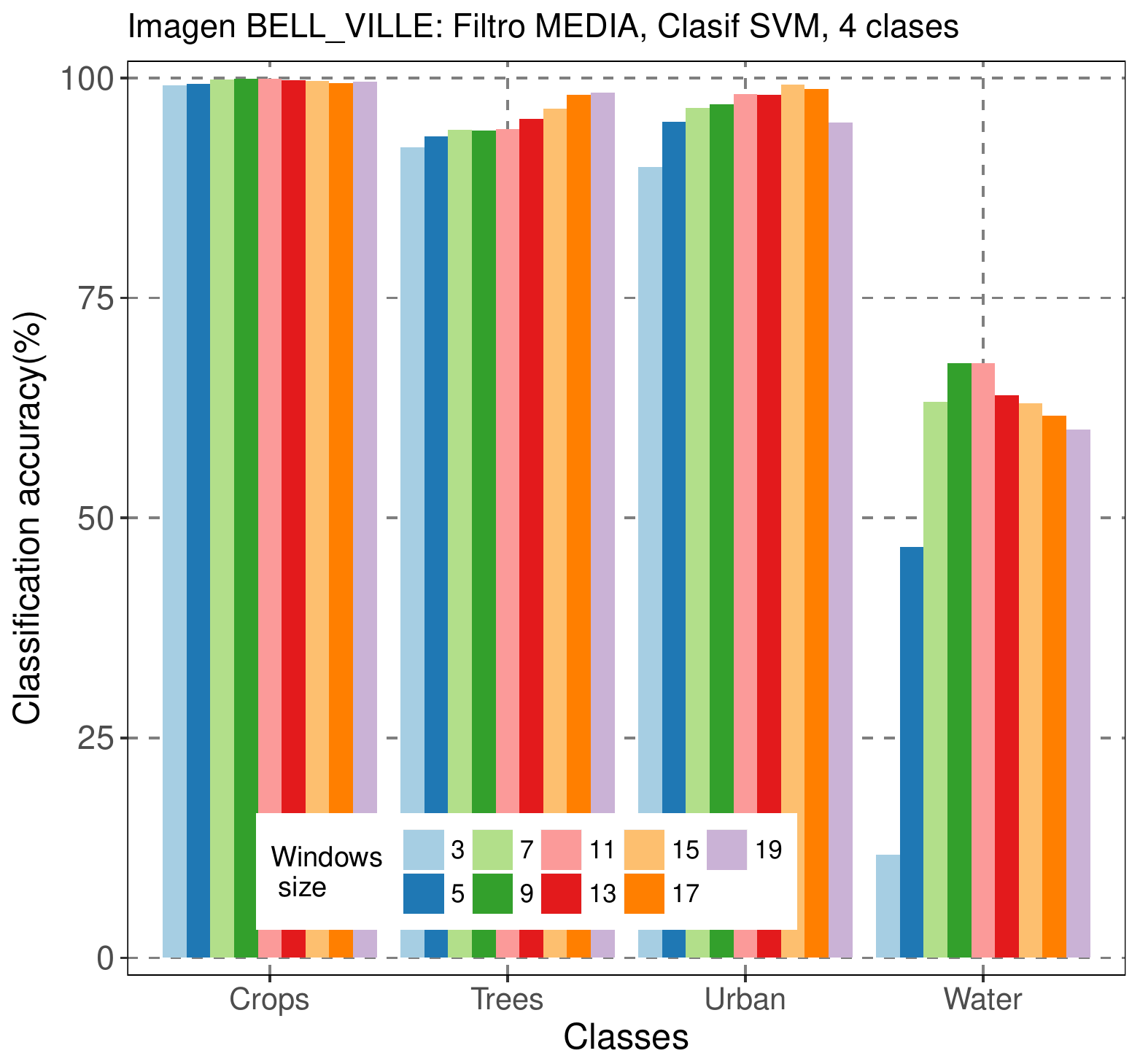}}
	\subfigure[Method 4 \label{porclase8}]{\includegraphics[width=.4\linewidth,trim ={0cm 0.1cm 0cm 0.7cm},clip,keepaspectratio]{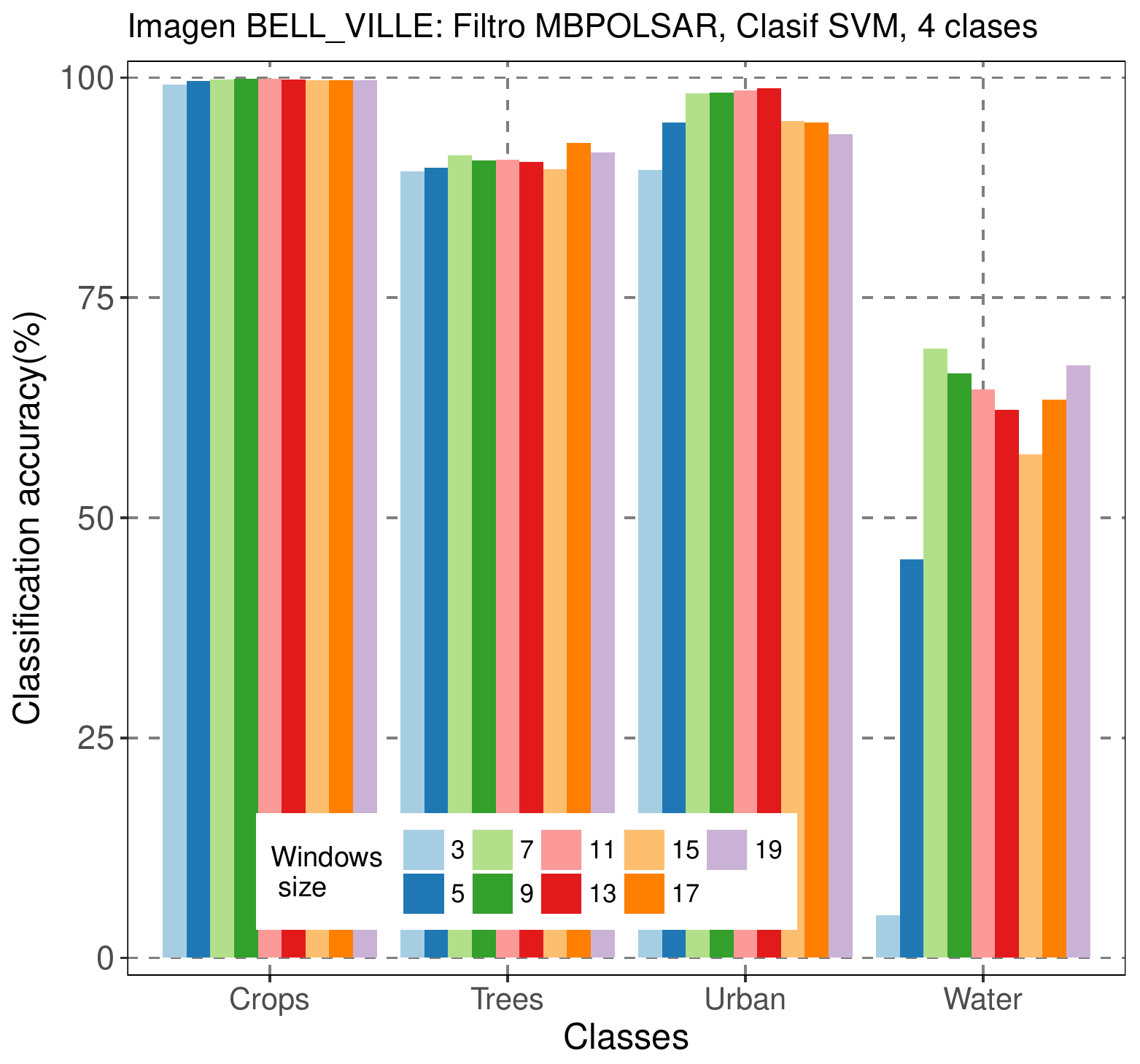}}
	\caption{UAVSAR data. Classification performance by class.\label{fig:porclaseBELL}}
\end{figure}

When the classes are considered individually, Fig.~\ref{fig:porclaseBELL}, the performances are similar when the same classifier is applied (as in San Francisco image).
Classification accuracy takes the maximum values for Crops class, and the minimum for Water class (river), specially using SVM where it is the class with worst performance with all degradations.

The classes Trees and Urban are very well classified with SVM and badly classified by ML (specially with small windows). 

\begin{figure}[htb]
	\centering
	\subfigure[Method 1 \label{clasif5}]{\includegraphics[width=.4\linewidth,trim ={0cm 3.7cm 0cm 3.3cm},clip,keepaspectratio]{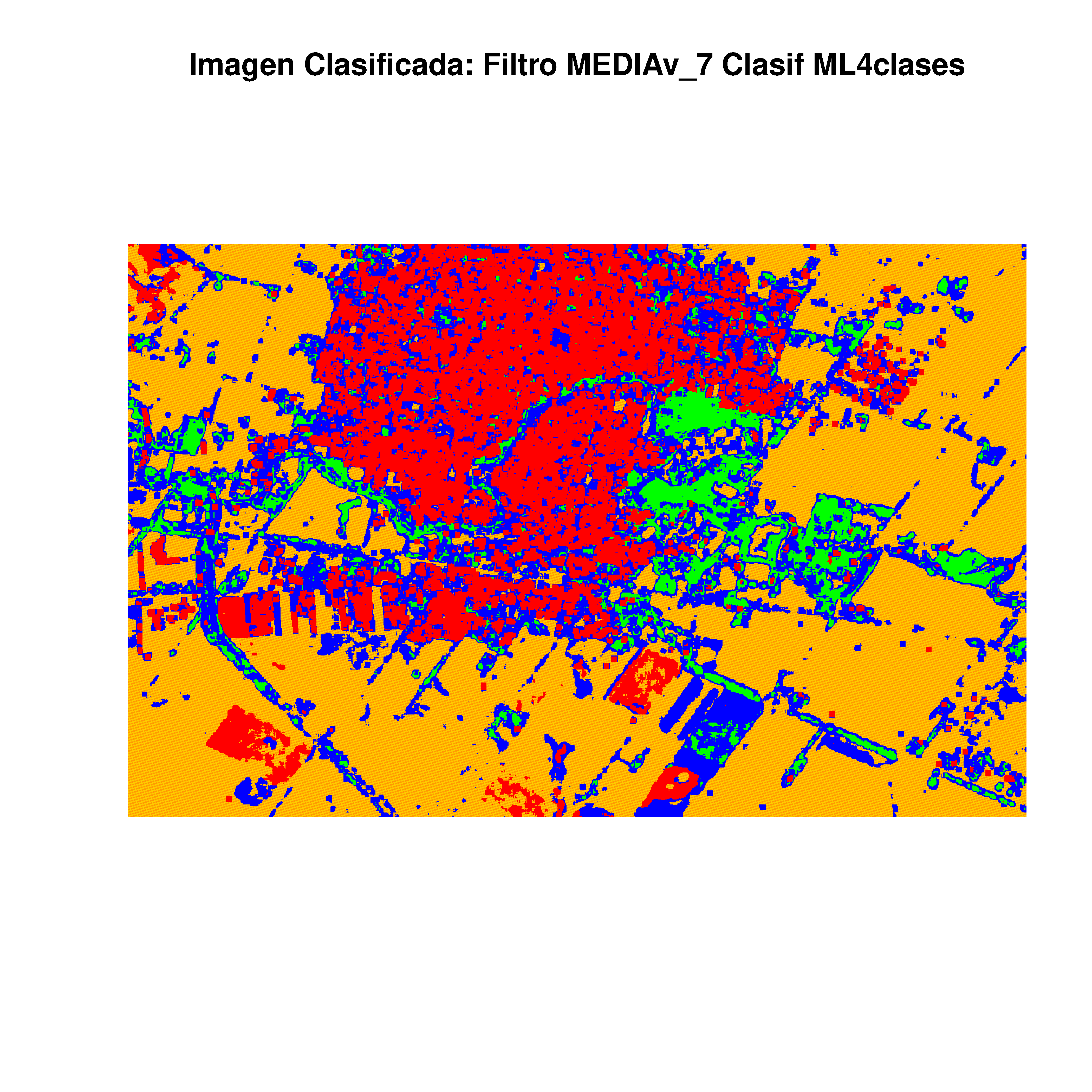}}
	\subfigure[Method 2 \label{clasif6}]{\includegraphics[width=.4\linewidth,trim ={0cm 3.7cm 0cm 3.3cm},clip,keepaspectratio]{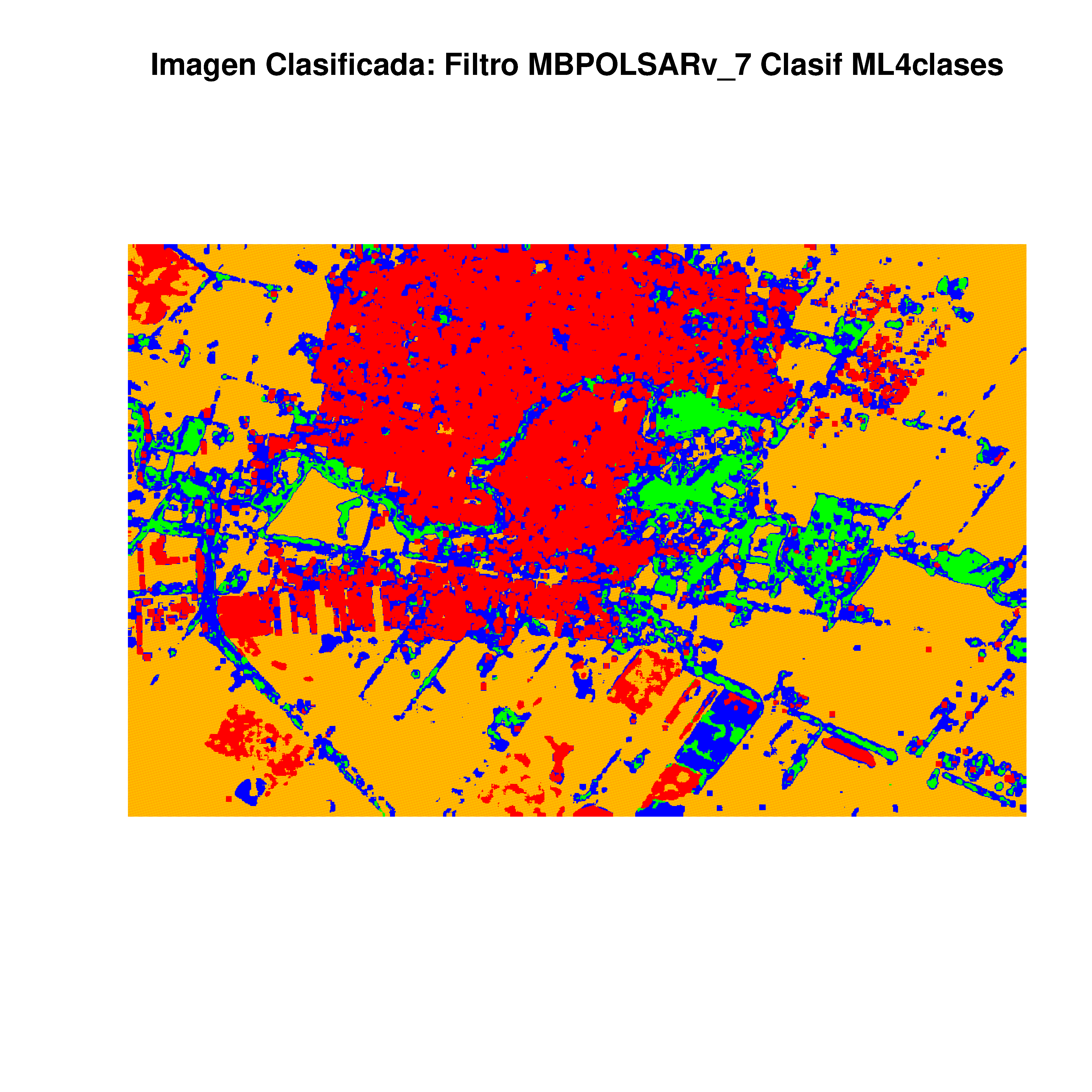}}
	\subfigure[Method 3 \label{clasif7}]{\includegraphics[width=.4\linewidth,trim ={0cm 3.7cm 0cm 3.3cm},clip,keepaspectratio]{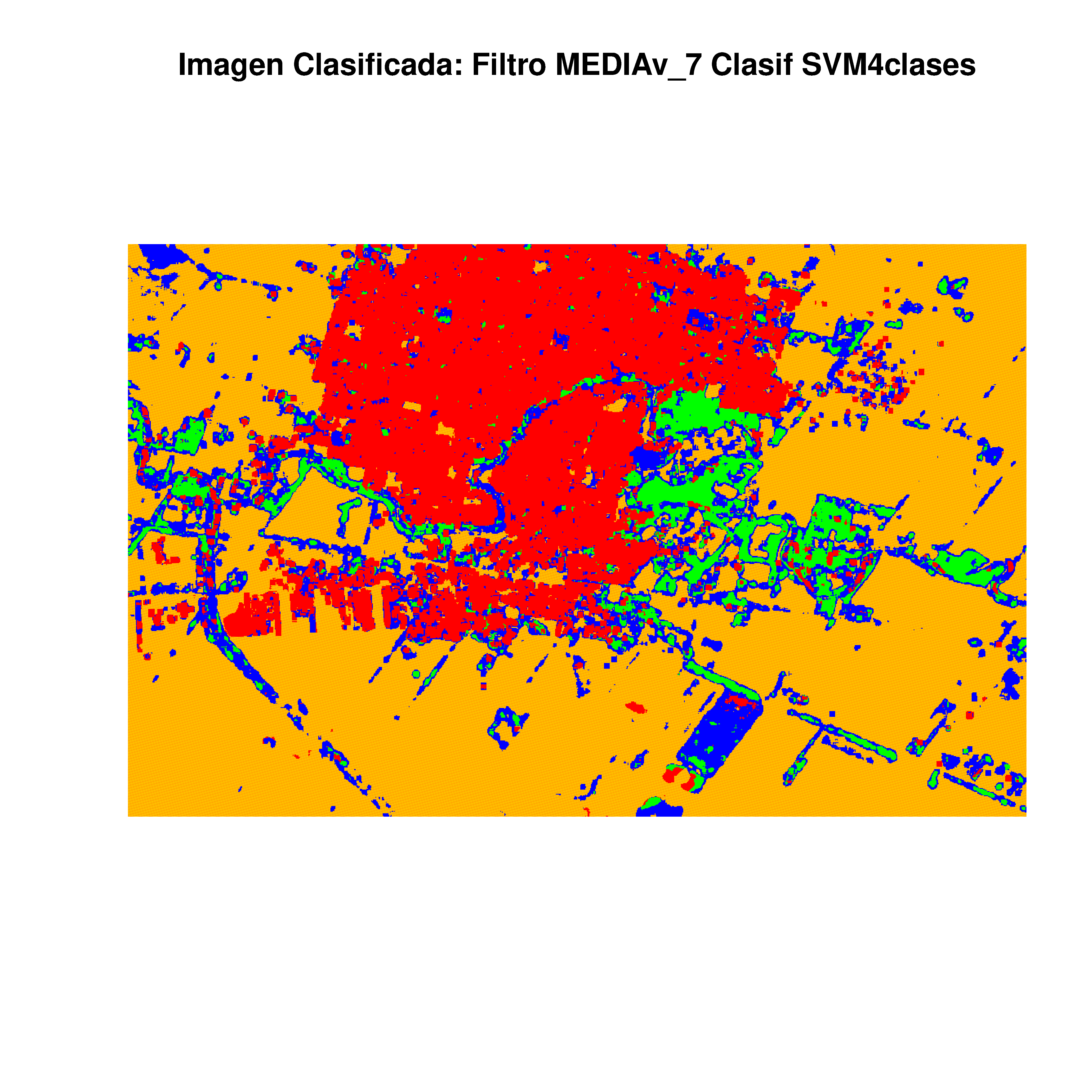}}
	\subfigure[Method 4 \label{clasif8}]{\includegraphics[width=.4\linewidth,trim ={0cm 3.7cm 0cm 3.3cm},clip,keepaspectratio]{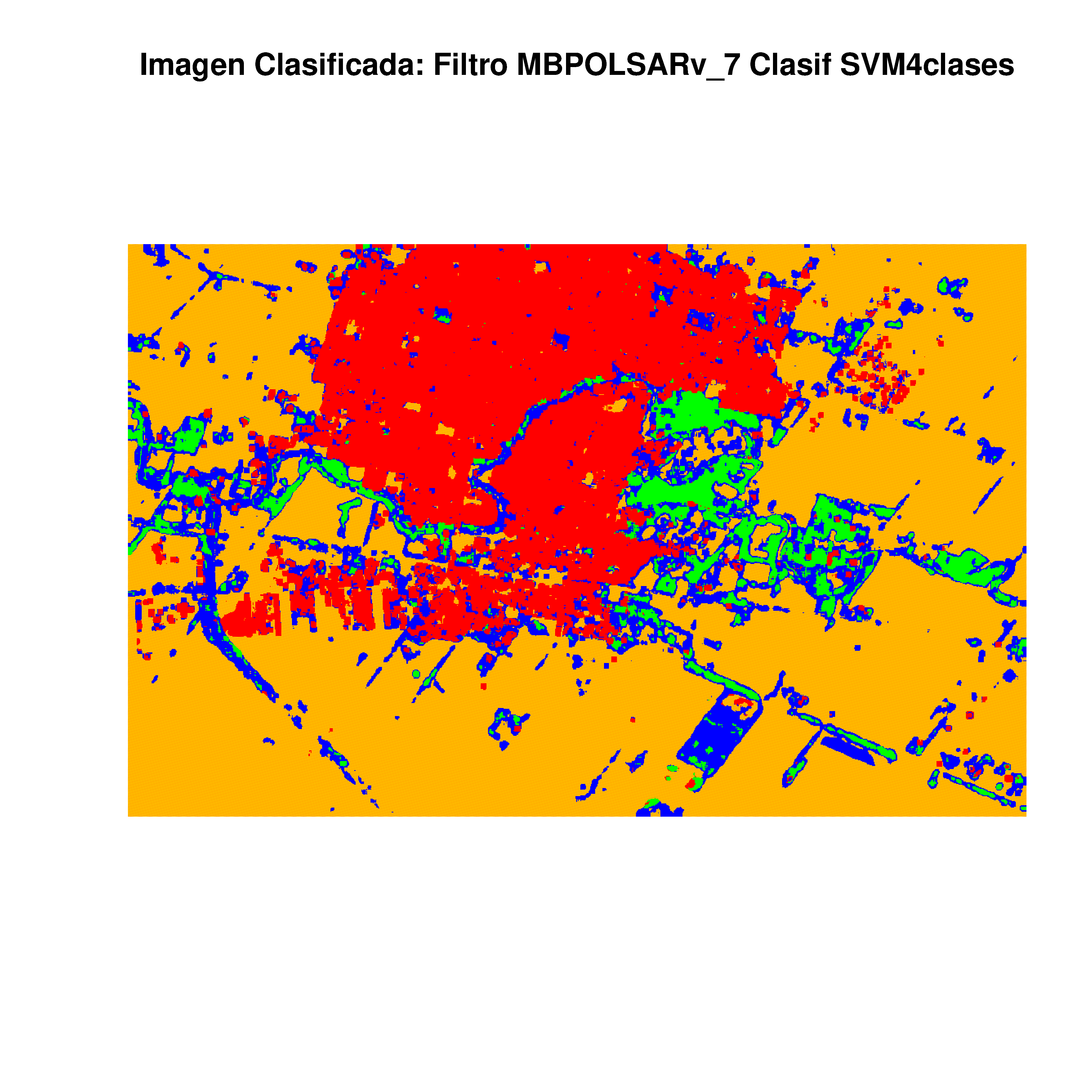}}
	\caption{UAVSAR data. Thematic maps using \num{7x7} window.\label{fig:mapasBELL}}
\end{figure}

The river is better classified by ML (although the values are just over  \SI{50}{\percent}) while with SVM and small windows the values 
are smaller which makes it invisible, Fig.~\ref{fig:mapasBELL}.
In addition, the Crops area is better conserved when we use SVM.

\section{CONCLUSION}
\label{sec:Conc}

As in~\cite{Narayanan02}, we studied how the information contents are affected by spatial resolution, and we compare the performances of four methods resulting from de combination of two filters (LM and MBPolSAR) and two classifiers (ML and SVM).

Globally the F and Kappa values are similar.

Almost all values are good (except for San Francisco image with Method $2$ and \num{3x3} window), and increase with degradation, i.e. filtering always improves the classification results at least up to \num{7x7}.

For San Francisco image the most noticeable difference between SVM and ML is that the first has the best performance with greater degradations whereas these windows spoil the ML classification.

When the classes are considered individually, the performances are similar when the same classifier is applied.

Not all the classes are best classified for the same method in the same situation, as the image may contain different amounts of information depending on the application~\citep{Narayanan02}.

\bibliographystyle{tfcad}
\bibliography{bibtesis}

\end{document}